\documentstyle[aps,bezier,12pt]{revtex}

\begin{document}
\tightenlines
\title{Copolymer Networks and Stars: Scaling Exponents}
\author{C.~von~Ferber$^{1,2}$ and Yu.~Holovatch$^3$}
\address{
$^1$School of Physics and Astronomy, Tel Aviv University
IL-69978 Tel Aviv, Israel \\
$^2$Fachbereich Physik, Universit\"at - Gesamthochschule - Essen,
D-45117 Essen, Germany \\
$^3$Institute for Condensed Matter Physics,
Ukrainian Academy of Sciences,
UA-290011 Lviv, Ukraine
}
\maketitle

\begin{abstract}
{
We explore and calculate the rich scaling behavior of copolymer
networks in solution by renormalization group methods.
We establish a field theoretic description in terms of composite
operators. Our 3rd order resummation of the spectrum of scaling
dimensions brings about remarkable features:
The special convexity properties of the spectra
allow for a multifractal interpretation while preserving stability
of the theory. This behavior could not be found
for power of field operators of usual $\phi^4$ field theory.
The 2D limit of the mutually avoiding walk star apparently
corresponds to results of a conformal Kac series. Such a classification
seems not possible for the 2D limit of other copolymer stars.
We furthermore provide a consistency check of two complementary
renormalization schemes: epsilon expansion and renormalization at
fixed dimension, calculating a large collection of independent
exponents in both approaches.
}
\end{abstract}

\section{Introduction}\label{I}
Recently, there has been considerable interest in the relation of
field theory and multifractals
\cite{Duplantier91,Ludwig90,Deutsch94,Janssen94}
and the associated multifractal dimension spectra
\cite{Cates87,Fourcade87,Chabra91,Halsey92,Halsey96,Ferber97}
as well as non-intersecting random walks and their 2D conformal theory
\cite{Duplantier88,Duplantier88a,Li90,Saleur92,Duplantier93}.
We present a model of multi-component polymer networks that shows a
common core of these topics and allows for a detailed study of the
interrelations.
While a multifractal spectrum may be derived from the scaling exponents of
two mutually avoiding stars of random walks
\cite{Duplantier91,Cates87}, a description in terms of power of field
operators seems to be ruled out by stability considerations
\cite{Duplantier91}. Here we show that already a simple product of two
power-of-field operators complies with both somewhat contrary
requirements.
We start from the theory of polymer stars, star - like arrangements of
polymer chains with self and mutual excluded interaction
\cite{Myake83,Duplantier86}.
We generalize this concept to stars of chains of different
polymer species which may differ in their self and mutual
interactions. Our formalism describes
homogeneous polymer stars, stars of mutually avoiding walks (MAW), and
the situation of two mutually interacting stars.

\begin{figure} [htbp]
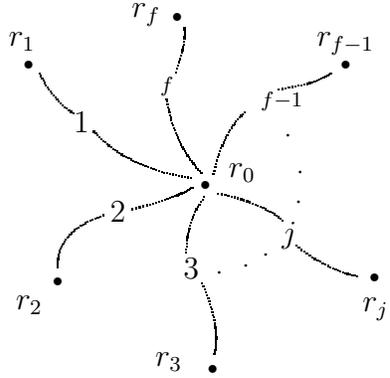

\begin{center}
\input fig1.pic
\end{center}
\caption{ Star polymer of $f$ arms linked together at point $r_0$
with extremities placed at points $r_1 \dots r_f$.
}
\label{fig1}
\end{figure}

\begin{figure} [htbp]
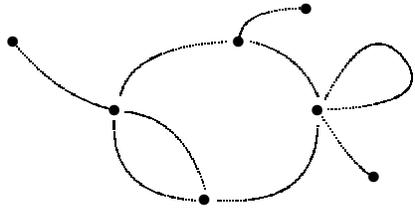

\begin{center}
\input fig2.pic
\end{center}
\caption{
A polymer network $\cal G$. It is characterized by
the numbers $n_{f}$ of
$f$-leg vertices. Here $n_{1}=3$, $n_{3}=2$,
$n_{4}=1$, $n_5=1$.
}
\label{fig2}
\end{figure}

\begin{figure} [htbp]
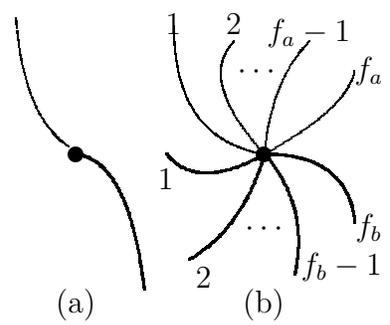

\begin{center}
\input fig3.pic
\end{center}
\caption{ $a.$ Block copolymer consisting of two polymer
chains of different species (shown by solid and thin lines) linked
at their endpoints.  $b.$ Copolymer star consisting of $f_a$ arms of
species $a$ and $f_b$ arms of species $b$ tied together at their
endpoints.  }
\label{fig3}
\end{figure}

Polymers and polymer solutions are among the most intensively studied
objects in condensed matter physics \cite{desCloizeaux90}.
The behavior of multi-component solutions containing polymers of different
species is especially rich.
Systems of chemically linked polymer chains of different species like
block copolymers are also of considerable experimental and
technical interest.  Linking polymer chains of different or even
contrary properties, like hydrophile and hydrophobe chains,  one
receives systems with qualitatively new behavior.  Here, we
concentrate on systems of chains in a solvent which, in a given
temperature range, differ  in their respective steric interaction
properties. We generalize the theory of multi-component polymer
solutions \cite{Joanny84,Schaefer85,Douglas86,Schaefer90,Douglas90} to
solutions of copolymer networks, i.e.  chains of different species
linked at their endpoints in the form of stars or networks of any
topology (see Figs.\ref{fig1} - \ref{fig3}).

For solutions of polymer networks of a single species the scaling
properties have been extensively studied by renormalization group
methods (for a review see \cite{Duplantier89}).
Star polymers as the most simple
polymer networks may be produced by linking together the endpoints of
polymer chains at some core molecule (Fig. \ref{fig1}).

In the same way networks of any given topology may be generated
(Fig. \ref{fig2}). Randomly linked polymer networks are also obtained
as a result of a vulcanization process by randomly linking nearby
monomers of different chains to each other.

The asymptotic properties of homogeneous systems of li\-ne\-ar cha\-in
mo\-le\-cu\-les in so\-lu\-ti\-on are universal in the limit of long
chains. Let us give a short account of the standard textbook results
\cite{desCloizeaux90,Yamakawa71,deGennes79,Freed86,Schaefer97}.
For each system one finds a so called $\Theta $
temperature at which the two point attractive and repulsive
interactions between the different monomers compensate each other.
As a result the polymer chains may be described by random walks (up to
higher order corrections):  The mean square distance between the
chains endpoints $\langle R^2\rangle$ scales with the number of
monomers $N$ like $\langle R^2\rangle\sim N$.  Above the $\Theta$
temperature the effective interaction between the monomers is
repulsive resulting in a swelling of the polymer coil which is
universal for $N\to\infty$:
\begin{equation}\label{1.1} \langle
R^{2}\rangle \sim N^{2\nu},
\end{equation}
where the correlation length exponent $\nu=0.588, \, 3/4$ in
space dimension $d=3, \, 2$.  The number of
configurations $\cal Z$ of a polymer chain of $N$ monomers grows for
$N\to \infty$ like
\begin{equation}\label{1.2}
{\cal Z} \sim e^{\mu N} N^{\gamma -1},
\end{equation}
with a non - universal fugacity $e^{\mu}$ and a universal exponent
$\gamma= 1.160, \, 43/32$ for $d=3, \, 2$.
In the early 70-ies following the work of de Gennes
\cite{deGennes72} the scaling theory of polymers was elaborated in
detail using the analogy between the asymptotic properties of long
polymer chains and the long distance correlations of a magnetic system
in the vicinity of the 2nd order phase transition
(see \cite{desCloizeaux90,deGennes79}).  This
mapping allows one to receive the above defined exponents $\nu$ and
$\gamma$ as limits of the correlation length exponent $\nu$
and the magnetic susceptibility critical exponent $\gamma$ of the
$O(m)$ - symmetric $m$-vector model in the formal limit $m\to 0$
\cite{deGennes72}.

On the other hand, if polymers of different species are present in the same
solution the scaling behavior of the observables may be much more rich.
Let us consider a solution of two different species of polymers in some
solvent, a so called ternary solution. Depending on the temperature the
system may then behave as if one or more of the inter- and intra- chain
interactions vanish in the sense described above
\cite{Joanny84,Schaefer85,Douglas86,Schaefer90,Douglas90}.
This will lead to asymptotic scaling laws that may differ from those observed
for each species alone \cite{Schaefer91}.

Interesting new systems are obtained when linking together
polymers of different species. The most simple system of this kind is
a so called block copolymer consisting of two parts of different
species.  They are of some technical importance e.g. serving as
surfactants \cite{Bates90}. For our study they give the
most simple example of a polymer star consisting of chains of two
different species (Fig. \ref{fig3}a) which we will call here a
copolymer star.  For the homogeneous polymer star the asymptotic
properties are uniquely determined by the number of its constituting
chains and the dimension of space .
For the number of configurations (partition function)
${\cal Z}_f$ of a polymer star of $f$
chains each consisting of $N$ monomers one finds
\cite{Duplantier86,Duplantier89}:
\begin{equation}\label{1.3}
{\cal Z}_f \sim e^{\mu N f} N^{\gamma_f -1}
\sim (R/\ell)^{\eta_f-f\eta_2}
\mbox{ , } N\to\infty  .
\end{equation}
The second part shows scaling with the size $R\sim N^\nu$ of the
isolated coil of $N$ monomers on some scale $\ell$, omitting the
fugacity factor.
The exponents $\gamma_f,\eta_f$, $f=1,2,3,\ldots$ constitute
families of `star exponents', which depend on the number of arms $f$
in a nontrivial way.  The case of linear polymer chains is included in
this family with the exponent $\gamma=\gamma_1=\gamma_2$  defined in
(\ref{1.2}). For general numbers of arms $f$ the star exponents
$\gamma_f,\eta_f$ have no physical counterparts in the set of
exponents describing magnetic phase transitions.  Nevertheless they
can be related to the scaling dimensions of composite operators of
traceless symmetry in the polymer limit $m\to 0$ of the $O(m)$
symmetric $m$-vector model \cite{Schaefer92,Wallace75}.  The
exponents $\gamma_f$ have been calculated analytically in perturbation
theory \cite{Myake83,Duplantier86,Schaefer92,Wallace75,Ohno88}, by
exact methods in two dimensions \cite{Duplantier86,Saleur86}, and by
Monte Carlo simulations  \cite{Barrett87,Batoulis89}.

It has been shown that the scaling properties of polymer networks
of arbitrary but fixed topology are uniquely defined by its constituting
stars \cite{Duplantier89}, as long as the statistical ensemble
respects some conditions on the chain length distribution
\cite{Schaefer92}.  Thus the knowledge of the set of star exponents
$\gamma_f$ or $\eta_f$
allows to obtain the power laws corresponding to (\ref{1.3}) also for
any polymer network of arbitrary topology.
The partition function ${\cal Z}_{\cal G}$ of a polymer network
$\cal G$ (see Fig.\ref{fig2}) of $F$ chains each of $N$ monomers
scales with $N\to\infty$ according to \cite{Duplantier86}:
\begin{equation} \label{1.3a}
{\cal Z}_{\cal G}\sim e^{\mu FN} N^{\nu(\eta_{\cal G}-F\eta_2)}
\mbox{ , with } \eta_{\cal G}=-dL+\sum_{f\geq 1} n_f\eta_f,
\end{equation}
where  $n_f$ the number of vertices with $f$ legs and $L=1+\sum (f/2-1)n_f$
is the number of loops in the network $\cal G$.

In this article we address a somewhat more complex problem:
What happens to the scaling laws if we build a polymer star or general
network of chains of different species?
In view of the above introduced ternary solutions, we thus study
systems of polymer networks in which some of the intra and inter chain
interactions may vanish.
For instance,there may be {\em only mutual} excluded volume
interactions between chains of the two different species in the
copolymer star shown in \ref{fig3} while chains of the same species
may freely intersect. Each subset of chains of one species thus
constitutes a star of random walks avoiding the second star of random
walks. Cates and Witten \cite{Cates87} have shown that this problem of
a star of random walks avoiding a given fractal structure is the key
to calculating the the multifractal spectrum of absorption of
diffusing particles on this fractal. This calculation can
be performed explicitly for the diffusion near absorbing polymers
\cite{Cates87,FerHol96d}.

The setup of our article is as follows. In section \ref{II} we introduce
notation and relate the polymer model to a Lagrangian field theory.
This field theoretical formalism will be used throughout the paper.
In section \ref{III} we define the renormalization group procedures.
We present two alternative approaches: zero mass renormalization together
with $\varepsilon$- expansion (see e.g.\cite{Brezin76}) and massive
renormalization at fixed dimension \cite{Parisi80}. Section \ref{IV}
is devoted to the study of the renormalization group flow of the
ternary model and its fixed points. Series for critical exponents governing
the scaling behavior of copolymer stars  and stars of mutually
avoiding walks are obtained in section \ref{V}.
In section \ref{VI} we discuss the problem of resummation of the
asymptotic series arising in this context. Numerical results are
presented in section \ref{VII}. We close with concluding remarks and
an outlook on possible applications of the theory in section
\ref{VIII} and give some calculational details in appendices.
Some of our main results we have announced in a letter
\cite{FerHol96e}.

\section{Model and Notations}
\label{II}

Let us first take a look at the model we use to describe polymers. In a first
discrete version we describe a configuration of the polymer by a set of
positions of segment endpoints:
\[ \mbox{Configuration} \{ r_1, \ldots , r_N \} \in I\!\!R^{d\times N} .
\]
Its statistical weight (Boltzmann factor) with the Hamiltonian
$\cal H$ divided by the product of Boltzmann constant $k_{\rm B}$ and
temperature $T$ will be given by
\begin{equation}\label{2.1}
\exp[-\frac{1}{k_{\rm B}T} {\cal H}]
=
\exp\{-\frac{1}{4\ell_0^2}\sum_{j=1}^{N}(r_j - r_{j-1})^2
  - \beta\ell_{0}^d \sum_{i\neq j=1}^N \delta^d(r_i - r_j) \}.
\end{equation}
The first term describes the chain connectivity, the parameter
$\ell_{0}$ governs
the mean sqare segment length. The second term describes the excluded
volume interaction forbidding two segment end points to take the same
position in space.  The parameter $\beta$ gives the strength of this
interaction.  The third parameter in our model is the chain length or
number of segments $N$.
The partition function $\cal Z$ is calculated as an integral over
all configurations of the polymer divided by the system volume
$\Omega$, thus dividing out identical configurations just translated
in space:
\begin{equation}\label{2.2}
{\cal Z}(N) = \frac{1}{\Omega}\int\prod_{i=1}^{N} {\rm d}r_i
\exp[-\frac{1} {k_{\rm B}T} {\cal H}\{r_i\}] .
\end{equation}
This will give us the `number of configurations' of the polymer
(\ref{1.2}). We will do our investigations by mapping the polymer
model to a renormalizable field theory making use of well developed
formalisms (see \cite{desCloizeaux90,deGennes79} for example).  To
this end we introduce a continuous version of our model as proposed by
Edwards \cite{Edwards65,Edwards66} generalizing it to describe a set
of $f$ polymer chains of varying composition possibly tied together at
their end points.  The configuration of one polymer is now given by a
path $r^a(s)$ in $d$ - dimensional space $I\!\!R^d$ parametrized by a
surface variable $0\leq s \leq S_a$. We now allow for a symmetric matrix
of excluded volume interactions $u_{ab}$ between chains
$a,b =  1,\ldots,f$. The Hamiltonian $\cal H$ is then given by
\begin{equation}
\label{2.3}
\frac{1}{k_{\rm B}T} {\cal H}(r^a) =
\sum_{a=1}^{f}\int_0^{S_a}{\rm d}s (\frac{{\rm
d}r^a(s)}{2{\rm d}s})^2 +
 \frac{1}{6}\sum_{a,b=1}^{f} u_{ab}\int{\rm
d}^d r \rho_a(r)\rho_b(r) ,
\end{equation}
with densities $\rho_a(r)=\int_0^{S_a}{\rm d}s
\delta^d(r-r^a(s))$ .
In this formalism the partition function is calculated as a
functional integral:
\begin{equation}
\label{2.4} {\cal Z}_f\{S_a\} =
\int{\cal D}[r^a(s)] \exp\{- \frac{1}{k_{\rm B}T}
{\cal H}(r^a)\} .
\end{equation}
here the symbol ${\cal D} [r_a(s)] $ includes normalization such that
$Z \{ S_a \} = 1$ for all $u_{ab}=0$. To make the exponential of
$\delta$-functions in (\ref{2.4}) and the functional
integral well-defined in the bare theory a cutoff
$s_0$ is introduced such that all simultaneous integrals of
any variables $s$ and $s^{ \prime }$
on the same chain are cut off by
$\mid s-s^{ \prime } \mid > s_0$. Let us note here that the
continuous chain model (\ref{2.3})
may  be understood as a limit of discrete  self-avoiding
walks, when the length of each step is decreasing $\ell_0\to 0$
while the number of
steps $N_a$ is increasing keeping the `Gaussian surface'
$S_a=N_a\ell_0^2$ fixed.
The continuous chain model (\ref{2.4}) can be mapped onto a
corresponding field theory by a Laplace transform in the Gaussian
surface varables $S_a$ to conjugate chemical potentials (``mass
variables'') $\mu_a$ \cite{Schaefer91}:
\begin{equation} \label{2.5}
\tilde{\cal Z}_f\{\mu_a\} = \int _0^{\infty} \prod_b {\bf d}S_b
e^{-\mu_b S_b} {\cal Z}_f\{ S_a \} .
\end{equation}
The Laplace-transformed partition function $\tilde{\cal
Z}_f\{\mu_a\}$ can be expressed as the $m = 0$ limit of the functional
integral over vector  fields $\phi_{a}, \, a=1,\ldots,f$ with $m$
components $\phi_a^{\alpha}, \, \alpha = 1,\ldots,m$ :
\begin{equation} \label{2.6}
\tilde{\cal Z}_f\{\mu_b\} =
  \int{\cal D}[\phi_a(r)]
   \exp[-{\cal L}\{\phi_b,\mu_b\}] \, |_{m=0}.
\end{equation}
The Landau-Ginzburg-Wilson Lagrangian $\cal L$ of $f$ interacting
fields $\phi_b$ each with $m$ components reads
\begin{equation}\label{2.7}
{\cal L}\{\phi_b,\mu_b\} = \frac{1}{2} \sum_{a=1}^{f}\int{\rm d}^d r
(\mu_a\phi_a^2 + (\nabla \phi_a(r) )^2 )
+ \frac{1}{4!} \sum_{a,a^{'}=1}^{f} u_{a,a'}
\int {\rm d}^d r \phi_a^2(r)\phi_{a'}^2(r) ,
\end{equation}
here
$  \phi_a^2 = \sum_{\alpha = 1}^{m}( \phi_a^{\alpha} )^2 $.
The limit $m=0$ in (\ref{2.6}) can be understood as a selection
rule for the diagrams contributing to the perturbation theory
expansions which can be easily checked diagrammatically. A formal
proof of (\ref{2.7}) using the
Stra\-to\-no\-vich-Hubbard transformation to linearize terms in
(\ref{2.3}) is given for the multi-component case in \cite{Schaefer91}.
The one particle irreducible vertex functions $\Gamma^{(L)}(q_i)$
of this theory are defined by:
\begin{equation}
\label{2.9}
\delta(\sum q_i) \Gamma^{(L)}_{a_1...a_L}(q_i) =
\int e^{iq_i r_i}
{\rm d} r_1 \dots  {\rm d} r_L
\langle \phi_{a_1}(r_1)\dots \phi_{a_L}(r_L)
\rangle^{\cal L}_{{\rm 1PI},m=0}.
\end{equation}
The average $\langle\cdots\rangle$ in (\ref{2.9}) is understood with
respect to the Lagrangian (\ref{2.7}) keeping only those
contributions which correspond to one-particle irreducible graphs
and which have non-vanishing tensor factors in the limit $m=0$.
The partition function $Z_{*f} \{ S_a \}$ of a  polymer  star
consisting of $f$ polymers of different species $1, \dots, f$
constrained to have a common end point  is obtained from
(\ref{2.4}) by introducing an appropriate product of
$\delta$-functions ensuring the  ``star-like'' structure. It reads:
\begin{eqnarray}
\label{2.10}
&& Z_{*f} \{ S_a \} =
\int {\cal D} [ r_a ]
\exp \{ -\frac{1}{k_{\rm B}T}{\cal H}(r_a)\}
\prod_{a=2}^f \delta^d(\vec{r}_a(0) - \vec{r}_1(0)).
\end{eqnarray}
The vertex part of its Laplace transformation may be defined by:
\begin{eqnarray}
\label{2.11}
&&\delta(p+\sum q_i)
\Gamma^{(*f)}(p,q_1 \dots q_f) =
\int e^{i(p r_0 + q_ir_i)}
{\rm d}^d r_0 {\rm d}^d r_1
\dots {\rm d}^d r_f
\nonumber \\ &&
 \langle \phi_1(r_0)\dots \phi_f(r_0)
\phi_1(r_1)\dots \phi_f(r_f)
\rangle^{\cal L}_{{\rm 1PI},{m=0}},
\end{eqnarray}
where all $a_1, \dots a_f$ are distinct.
The vertex function $\Gamma^{(*f)}$ is thus defined by insertion of
the composite operator $\prod_a\phi_a$. Its scaling properties
are governed by the scaling dimension of this operator.  When only one
species is present one can also define $\Gamma^{*f}$ by insertion
of a composite operator of traceless symmetry
\cite{Wallace75}.  In the following  we will be mainly interested in
the case of only two species of polymers,
with interactions $u_{11}$, $u_{22}$ between the polymers of the same
species and $u_{12} \, = \, u_{21}$  between the polymers of different
species. In this case the composite operator in (\ref{2.11}) reduces
to the product of two power-of-field operators with appropriate
symmetry $(\phi)^{f_1} (\phi^{\prime})^{f_2}$ each corresponding to a
product of fields of the same `species'.
Nevertheless, our results are easily generalized to the
case of any number of polymer species.

The starting point for our calculations are the three loop expansions
for the bare vertex functions of interest
$(\partial/\partial k^2) \Gamma^{(2)}$, $\Gamma^{(4)}$,
$\Gamma^{(*f)}$.
They involve the loop integrals $D_2, I_1 - I_8$. These are given in
appendix A together with their corresponding graphs.
The expressions read:
\begin{equation}
\label{2.13} \frac{\partial}{\partial k^2} \Gamma^{(2)}_{(aa)} = 1 -
\frac{1}{9}I_2 u_{aa}^2 + \frac{4}{27}I_8u_{aa}^3,
\end{equation}
\begin{eqnarray}
\label{2.14} \Gamma^{(4)}_{(aaaa)} &=& u_{aa} - \frac{4}{3}D^{aa}_{2}
u_{aa}^2 + (\frac{5}{9} D_2^2 + \frac{22}{9} I_1)u_{aa}^3 -
(\frac{2}{9} D_2^3 + \frac{28}{27} I_1D_2 + \nonumber \\
&& \frac{8}{27} I_3 + \frac{40}{9} I_4 +
 \frac{58}{27} I_5 + \frac{14}{27} I_6 + \frac{22}{27} I_7)u_{aa}^4,
\end{eqnarray}
\begin{eqnarray} \lefteqn{
\Gamma^{(*f)} = 1
+D^{a_1a_2}_{{2}}\bar{u}_{{a_{{1}}a_{{2}}}} /2
+{D_{{2}}^{2}}\bar{u}_{{a_{{1}}a_{{2}}}}\bar{u}_{{a_{{3}}a_{{4}}}} /8
+\bar{u}_{{a_{{1}}a_{{2}}}}\bar{u}_{{a_{{1}}a_{{3}}}}I_{{1}}
+\bar{u}_{{a_{{1}}a_{{1}}}}\bar{u}_{{a_{{1}}a_{{2}}}}I_{{1}}
}
\nonumber\\&&
+\left (I_{{1}}+{D_{{2}}^{2}}\right ){\bar{u}_{{a_{{1}}a_{{2}}}}^{2}}
/2
+{D_{{2}}^{3}}\bar{u}_{{a_{{1}}a_{{2}}}}\bar{u}_{{a_{{3}}a_{{4}}}}
\bar{u}_{{a_{{5}}a_{{6}}}}
/48
\nonumber\\&&
+D_{{2}}\bar{u}_{{a_{{1}}a_{{2}}}}\bar{u}_{{a_{{1}}a_{{3}}}}
\bar{u}_{{a_{{4}}a_{{5}}}}I_{{1}}
/2
+\bar{u}_{{a_{{1}}a_{{2}}}}\bar{u}_{{a_{{1}}a_{{3}}}}
\bar{u}_{{a_{{3}}a_{{4}}}}I_{{4}}
\nonumber\\&&
+\left (I_{{5}}+I_{{6}}\right )\bar{u}_{{a_{{1}}a_{{2}}}}
\bar{u}_{{a_{{1}}a_{{3}}
}}\bar{u}_{{a_{{2}}a_{{4}}}}
/2
+\bar{u}_{{a_{{1}}a_{{2}}}}\bar{u}_{{a_{{1}}a_{{3}}}}
\bar{u}_{{a_{{1}}a_{{4}}}}I_{{4}}
\nonumber\\&&
+\left (3\ I_{{4}}+I_{{7}}\right )\bar{u}_{{a_{{1}}a_{{2}}}}
\bar{u}_{{a_{{1}}a_{{
3}}}}\bar{u}_{{a_{{2}}a_{{3}}}}
/3
+D_{{2}}\left (I_{{1}}+{D_{{2}}^{2}}\right )
{\bar{u}_{{a_{{1}}a_{{2}}}}^{2}}\bar{u}_{{a_{{3}}a_{{4}}}}
/4
\nonumber\\&&
+\left (D_{{2}}I_{{1}}+2\ I_{{4}}+2\ I_{{5}}+I_{{7}}\right )
{\bar{u}_{{a_{{1}}a_{{2}}}}^{2}}\bar{u}_{{a_{{1}}a_{{3}}}}
+\left (I_{{3}}+3\ I_{{4}}+I_{{5}}\right )
{\bar{u}_{{a_{{1}}a_{{1}}}}^{2}}\bar{u}_
{{a_{{1}}a_{{2}}}}
\nonumber\\&&
+\left (D_{{2}}I_{{1}}+2\ I_{{4}}+I_{{5}}+I_{{6}}+{D_{{2}}^{3}}\right )
{\bar{u}_{{a_{{1}}a_{{2}}}}^{3}}
/2
+D_{{2}}\bar{u}_{{a_{{1}}a_{{1}}}}\bar{u}_{{a_{{1}}a_{{2}}}}
\bar{u}_{{a_{{3}}a_{{4}}}}
I_{{1}}
/2
\nonumber\\&&
+\left (I_{{4}}+I_{{5}}\right )\bar{u}_{{a_{{1}}a_{{1}}}}
\bar{u}_{{a_{{1}}a_{{2}}
}}\bar{u}_{{a_{{2}}a_{{3}}}}
+\left (I_{{4}}+I_{{5}}+I_{{7}}\right )\bar{u}_{{a_{{1}}a_{{1}}}}
\bar{u}_{{a_{{1}}
a_{{2}}}}\bar{u}_{{a_{{1}}a_{{3}}}}
\nonumber\\&&
+\bar{u}_{{a_{{1}}a_{{1}}}}\bar{u}_{{a_{{1}}a_{{2}}}}
\bar{u}_{{a_{{2}}a_{{2}}}}I_{{5}}
/2
+\left (D_{{2}}I_{{1}}+4\ I_{{4}}+I_{{6}}\right )
\bar{u}_{{a_{{1}}a_{{1}}}}{\bar{u}_{{a_{{1}}a_{{2}}}}^{2}}.
\label{2.15}
\end{eqnarray}
In (\ref{2.15}) summation over $a_i=1\ldots f$ is assumed.
The equations (\ref{2.13}-\ref{2.15}) apply to any number of polymer
species. For a star of $f_1$ chains of species $1$ and $f_2$ chains of
species $2$ we restrict $a=1,2$ in (\ref{2.13}),(\ref{2.14}) and
 the matrix of interactions $\bar{u}_{ab}$  is given by $$
 \bar{u}^{f_1f_2}_{ab}=\left\{\begin{array}{ll} u_{11}& 1 \leq a,b\leq
f_1\\ u_{22}& f_1 < a,b \leq f\\
u_{12}& \mbox{else}. \end{array} \right.
$$
Let us define in this way:
\begin{equation}
\label{2.16}
\Gamma^{(*f_1f_2)}=\Gamma^{(*f)}|_{\bar{u}_{ab} =
\bar{u}^{f_1f_2}_{ab}}.
\end{equation}
For general $f_1$, $f_2$ the corresponding combinatorics may also be
directly calculated by summation over $a_i=1,2$ instead.
Replacing $\bar{u}_{a_ia_j}=u_{a_ia_j}$, each term in the sum with
indices $a_1\dots a_k $ then acquires a factor
$$ \Big ( {{f_1} \atop {\#_1(a_1 \dots a_k)}} \Big ) \Big ( {{f_2} \atop
{\#_2(a_1 \dots a_k)}} \Big ).  $$
Here $\#_1(a_1 \dots a_k)$ is the number of $a_i=1$ whereas $\#_2(a_1
\dots a_k)$ is the number of $a_i=2$.

As a special case we may derive the vertex function
$ \Gamma^{(4)}_{1122} $ for the $u_{12}$
interaction  using the relation
$\Gamma^{(*22)}=\partial / \partial u_{12} \Gamma^{(4)}_{1122}$ which
is obvious from the perturbation theory (see \cite{Schaefer91} for
instance):
$$
\Gamma^{(4)}_{1122} = \int{\rm d} u_{12} \Gamma^{(*22)}.
$$
Note, that the vertex function $\Gamma^{(*20)}$ and $\Gamma^{(*11)}$
define a vertex function with a $\phi^2$ insertion which in standard
literature is denoted by $\Gamma^{(2,1)}$ \cite{Brezin76}.
With the same formalism we can also describe a star of $f$ mutually
avoiding walks \cite{Duplantier88,Duplantier88a}. In this case all
interactions on the same
chain $\bar{u}_{aa}$ vanish and only those $\bar{u}_{ab}$ with $a \neq
b$ remain:
\begin{equation} \label{2.17} \Gamma^{(*f)}_{\rm
MAW}=\Gamma^{(*f)}|_{\bar{u}_{ab}=(1-\delta_{ab})u_{12}}.
\end{equation}
In this case each term with indices ${a_1 \dots a_k}$
acquires a factor $\Big ({f \atop k} \Big ) k!$.

As is well known, ultraviolet divergences occur when the vertex
functions (\ref{2.13}) - (\ref{2.15}) are evaluated naively
\cite{Bogoliubov59}. In the next section we apply the field
theoretical renormalization group approach to remove the divergences
and to make transparent the scaling symmetry of the problem.

\section{Renormalization}\label{III}
We apply renormalization group (RG)
theory to make use of the scaling symmetry of
the system in the asymptotic limit to extract the universal content and
at the same time remove divergences which occur for the evaluation of
the bare functions in this limit \cite{Brezin76,Bogoliubov59,Amit84}.
The theory given in terms of the initial bare variables is mapped to a
renormalized theory. This is achieved by a controlled rearrangement of
the series for the vertex functions. Several asymptotically equivalent
procedures serve to this purpose. Here we will use two somewhat complementary
approaches: zero mass renormalization (see \cite{Brezin76} for instance)
with successive $\varepsilon$-expansion \cite{Wilson72}
and the fixed dimension massive RG approach \cite{Parisi80}.
The first approach is performed directly for the critical point .
Results for critical exponents at physically interesting
dimensions $d=2$ and $d=3$ are calculated in an $\epsilon=4-d$
expansion \cite{Wilson72,Brezin73,Vladimirov79,Gorishny84}.  The
second approach renormalizes off the critical limit but calculates
the critical exponents directly in space dimensions $d=2$, $d=3$
\cite{Nickel78,LeGuillou80}.
It also gives quantitative results for the preasymptotic
critical behavior \cite{Bagnuls85,Bagnuls87}. Most authors tend to
prefer one method and to exclude the other for non obvious reasons.
The application of both approaches will enable us in particular to
check the consistency of approximations and the accuracy of the
results obtained.

Let us formulate the relations for a renormalized theory in terms of
the corresponding renormalization conditions.
Though they are different in principle for the two procedures,
we may formulate them simultaneously using the same expressions.
Note that the polymer limit of zero component fields leads to essential
simplification. Each field $\phi_a$, mass $m_a$ and coupling $u_{aa}$
renormalizes as if the other fields were absent.
First we introduce renormalized couplings $g_{ab}$ by:
\begin{eqnarray}\label{3.1}
u_{aa} &=& \mu^{\varepsilon} Z^{-2}_{\phi_a}Z_{aa} g_{aa},
\hspace{1em} a=1,2
\\ \label{3.2}
u_{12} &=& \mu^{\varepsilon} Z_{\phi_1}^{-1}Z_{\phi_2}^{-1}Z_{12}
g_{12} .
\end{eqnarray}
Here, $\mu$ is a scale parameter equal to the renormalized
mass at which the massive scheme is evaluated and sets the scale
of the external momenta in the massless scheme. The renormalization
factors $Z_{\phi_a}, Z_{ab}$ are defined as power series in the
renormalized couplings which fulfill the following RG conditions:
\begin{eqnarray}
Z_{\phi_a}(g_{aa}) \frac{\partial}{\partial k^2}
\Gamma_{aa}^{(2)}(u_{aa}(g_{aa})) = 1,                  \label{3.3}\\
Z^2_{\phi_a}(g_{aa})
\Gamma_{aaaa}^{(4)}(u_{aa}(g_{aa})) = \mu^\varepsilon g_{aa},
\label{3.4}\\
Z_{\phi_1}(g_{11}) Z_{\phi_2}(g_{22})
\Gamma_{1122}^{(4)}(u_{ab}(g_{ab})) = \mu^\varepsilon g_{12}.
\label{3.5}
\end{eqnarray}
These formulas are applied perturbatively while the corresponding loop
integrals are evaluated for zero external momenta in the massive
approach and for external momenta at the scale of $\mu$ in the
massless approach as explained in appendix \ref{A}. In the massive
case the RG condition for the vertex function $\Gamma^{(2)}$ reads
\begin{equation}\label{3.6}
Z_{\phi_a}(g_{aa})
\Gamma_{aa}^{(2)}(u_{aa}(g_{aa}))|_{k^2=0}= \mu^2, \hspace{1em} a=1,2.
\end{equation}
In the case of massless renormalization the corresponding condition
reads \cite{Brezin76}:
\begin{equation}\label{3.7}
Z_{\phi_a}(g_{aa})
\Gamma_{aa}^{(2)}(u_{aa}(g_{aa}))|_{k^2=0} = 0, \hspace{1em} a=1,2.
\end{equation}
In order to renormalize the star vertex functions we introduce
renormalization factors $Z_{* f_1,f_2}$ by
\begin{equation}\label{3.8}
Z_{\phi_1}^{f_1/2}Z_{\phi_2}^{f_2/2}Z_{* f_1,f_2}
\Gamma^{(* f_1 f_2)}(u_{ab}(g_{ab})) = \mu^{\delta_{f_1+ f_2}}.
\end{equation}
In the same way we define the appropriate renormalization for the
vertex function of mutually avoiding walks (MAW):
\begin{equation}\label{3.9}
Z_{\phi_1}^{f/2}Z_{({\rm MAW} f)}
\Gamma_{{\rm MAW}}^{*f}(u_{12}(g_{ab})) = \mu^{\delta_f}.
\end{equation}
The powers of $\mu$ absorb the engineering dimensions of the bare vertex
functions. These are given by
\begin{equation}\label{3.10}
\delta_f =  f(\varepsilon/2 -1)+ 4 - \ \varepsilon.
\end{equation}
The renormalized couplings $g_{ab}$ defined by the relations
(\ref{3.1}),(\ref{3.2}) depend on the scale parameter $\mu$.
By their dependence on $g_{ab}$ also the renormalization $Z$ - factors
implicitly depend on $\mu$. This dependence
is expressed by the RG functions defined by the following
relations:
\begin{eqnarray}
\mu \frac{\rm d}{{\rm d}\mu} g_{ab} &=& \beta_{ab}(g_{a'b'}).
\label{3.12}\\
\mu \frac{\rm d}{{\rm d}\mu} \ln Z_{\phi_a} &=&
\eta_{\phi_a}(g_{aa}).                        \label{3.13}\\
\mu \frac{\rm d}{{\rm d}\mu} \ln Z_{*f_1f_2} &=&
\eta_{*f_1f_2}(g_{ab}).                        \label{3.14}\\
\mu \frac{\rm d}{{\rm d}\mu} \ln Z_{{\rm MAW}f} &=&
\eta^{{\rm MAW}}_f(g_{ab}).                        \label{3.15}
\end{eqnarray}
The function $\eta_{\phi_a}$ describes the pair correlation critical exponent,
while the functions $\eta_{*f_1f_2}$ and $\eta^{{\rm MAW}}_f(g_{ab})$
define the set of exponents for copolymer stars and stars of mutually
avoiding walks. Note that $Z_{*20}$ renormalizes the vertex function
with a $\phi^2$ insertion which coincides with $\Gamma^{(*20)}$.
Consequently, the usually defined correlation length critical
exponent $\nu$ is expressed in terms of functions  $\eta_{*20}$ and
$\eta_{\phi}$ (see next chapter). Explicit expressions for the $\beta$
and $\eta$ functions will be given in the next section together with a
study of the RG flow and the fixed points of the theory.

\section{
Renormalization Group Flow and the
Fixed Points: $\varepsilon$-expansion and
pseudo-$\varepsilon$-expansion
}
\label{IV}
Here, we want to discuss the RG flow of the theory presented in
section \ref{III}. In particular, we want to find appropriate
representations for the fixed points of the flow.
In a study devoted to ternary polymer solutions, the RG flow has
been calculated \cite{Schaefer91} within massless
renormalization and is known to the third loop order of the
$\varepsilon$-expansion. Note, that for the diagonal coupling $g_{aa}$
the corresponding expressions are also found in the polymer limit
$m=0$ of the $O(m)$-symmetric $\phi^4$ model. They are known in even
higher orders of perturbation theory \cite{note2}.
To third loop order the expressions read:
\begin{eqnarray} &&
\beta_{g_{aa}}^{\varepsilon} = -\varepsilon g_{aa} +
 \frac{1}{3}(4 + 2 \varepsilon + 2 \varepsilon^2) g_{aa}^2 -
\frac{1}{9} (\frac{21}{2} + \frac{215}{8} \varepsilon -
11 J \varepsilon) g_{aa}^3 +
\nonumber \\ &&
\frac{1}{27} (79 - 22 J + 33 \zeta(3)) g_{aa}^4 +
O(g_{aa}^5), \hspace{3em} a=1,2.
\label{4.1}
\\ &&
\beta_{g_{12}}^{\varepsilon}  = -\varepsilon g_{12} +
\frac{1}{3}(1 + \frac {\varepsilon}{2} +
\frac {\varepsilon^2} {2})  (g_{11} + g_{22}) g_{12}
+ \frac{1}{3}(2 +
\varepsilon + \varepsilon^2) g_{12}^2 -
\nonumber \\ &&
\frac{1}{9} (\frac{5}{4} +
\frac{55}{16} \varepsilon -
\frac{3}{2} J \varepsilon) (g_{11}^2 + g_{22}^2)g_{12} -
\frac{1}{9} (3 + \frac{15}{2} \varepsilon -
3 J \varepsilon) (g_{11} + g_{22})g_{12}^2 -
\nonumber \\ &&
\frac{1}{9} (2 +
5 \varepsilon - 2 J \varepsilon) g_{12}^3 +
\frac{1}{54}(15 - J)(g_{11}^3 + g_{22}^3)g_{12} +
\frac{1}{27}(\frac{27}{2} +
\nonumber \\ &&
9 \zeta(3) - 6 J)(g_{11}^2 +
g_{22}^2)g_{12}^2 +
\frac{1}{27}(7 - 3 J)g_{11} g_{22} g_{12}^2 +
\frac{1}{27}(12 + 6 \zeta(3) -
\nonumber \\ &&
2 J)(g_{11} + g_{22})g_{12}^3 +
\frac{1}{27}(6 + 3 \zeta(3) - 2 J) g_{12}^4 +
O(g^5).
\label{4.2}
\end{eqnarray}
Here, the Riemann $\zeta$-function with $\zeta(3) \approx 1.202$ and
the constant $J \approx 0.7494$ occur. We use an index
$\varepsilon$ at $\beta^{\varepsilon}$ to distinguish the
$\beta$-functions obtained in massless renormalization with
successive $\varepsilon$-expansion from $\beta^m$ obtained in massive
field theory.

Similarly, performing renormalization in the massive scheme,
we obtain  the corresponding functions
$\beta^m$ . We present them using convenient variables
$v_{ab}=D_2^m g_{ab}$ and introduce new functions
$\beta^m_{v_{ab}}=D_2^m \beta^m_{g_{ab}}$. Here, $D_2^m$ is the
one-loop integral calculated within massive field theory  (see
appendix A). This procedure defines a convenient numerical scale for
the massive $\beta$-functions. The expressions for the functions
$\beta^m_{v_{ab}}$ read:
\begin{eqnarray} \nonumber &&
\beta_{v_{aa}}^m = - (4-d) v_{aa} ( 1 - \frac{4v_{aa}}{3} +
\frac{2}{9} ( 22 (i_1 - \frac{1}{2})+ 2  i_2  ) v_{aa}^2 +
\frac{2}{27} (- 89+ \\ &&  \nonumber
310 i_1 + 8 i_2 + 3  i_2 d-
12  i_3 - 180  i_4 - 87 i_5 - 21  i_6 -
33 i_7 - \\ &&
12 i_8  ) v_{aa}^3 )
+O(v^5)
, \hspace{5.0em} a=1,2. \hspace{0.5em}
\label{4.3} \\ &&
\beta_{v_{12}}^m = - (4-d) v_{12}  ( 1 - \frac{1}{3}   (v_{11} +
v_{22} + 2v_{12}   ) + \frac{1}{3}   (- v_{11}^2 -
v_{22}^{2} - 2v_{12}v_{11} \nonumber \\ &&
-\frac{4v_{12}^{2}}{3} - 2v_{12}v_{22} +
\frac{2v_{12}^{2} i_2}{3}+\frac {8v_{12}^{2} i_1}{3}+ 4v_{12} v_{22}
i_1 + 2v_{11}^2 i_1+
\frac {2v_{22}^2 i_2}{3}+ \nonumber \\ &&
4v_{12}v_{11} i_1 + 2v_{22}^2 i_1   )
+ \sum_{jkl} b^{jkl}v_{11}^j v_{22}^k v_{12}^l   ) +O(v^5).
\label{4.4}
\end{eqnarray}
Here, $i_j$ are the dimension-dependent loop integrals,
normalized by the one-loop integral value (see appendix \ref{A}).
Expressions for the coefficients $b^{jkl}$ are given in the
appendix \ref{B}.
Note that the $\beta$-function for the diagonal coupling $g_{aa}$ is
known within the massive scheme \cite{Nickel78} to the order of six loops
\cite{note3}.

\begin{figure} [htbp]
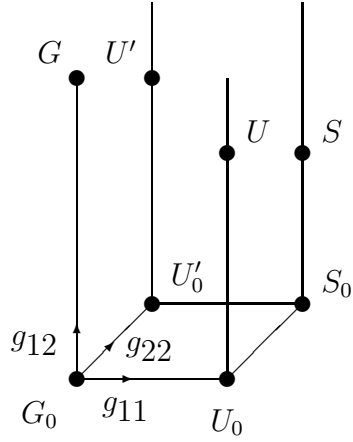

\begin{center}
\input fig4.pic
\end{center}
\caption{ Fixed points (FPs) of ternary polymer solution.
The trivial FPs $G_0$, $U_0$, $U_0^{\prime}$, $S_0$ correspond to
vanishing mutual interaction. The non-trivial FPs $G$, $U$,
$U^{\prime}$, $S$ correspond to non-vanishing mutual interaction
($g_{12} \neq 0$).  } \label{fig4}
\end{figure}

Let us solve the equations for the fixed points (FP) P$(\{g^*_{11},
g^*_{22}, g^*_{12}\})$ of the $\beta$-functions,
\begin{eqnarray}
\beta_{g_{aa}}^{\varepsilon} (g^*_{aa})&=& 0, \hspace{2em} a=1,2,
\nonumber \\
\beta_{g_{12}}^{\varepsilon}(g^*_{11}, g^*_{22}, g^*_{12}) &=& 0.
\label{4.5}
\end{eqnarray}
As is well known, the first equation has two solutions
$g^*_{aa}=0,g^*_{\rm S}$. For the second one finds a total of 8 FP
depending on the choice of $g^*_{aa}$.
The trivial FPs are ${\rm G}_0(0,0,0), {\rm U}_0(g^*_{\rm S},0,0),
{\rm U'}_0(0,g^*_{\rm S},0), {\rm S}_0(g^*_{\rm S},g^*_{\rm S},0)$,
all corresponding to $g^*_{12}=0$.
The non trivial FPs are found as
${\rm G}(0,0,g^*_{\rm G})$,
${\rm U}(g^*_{\rm S},0,g^*_{\rm U})$,
${\rm U'}(0,g^*_{\rm S},g^*_{\rm U})$, and
${\rm S}(g^*_{\rm S},g^*_{\rm S},g^*_{\rm S})$.
In the three dimensional space of couplings $g_{11}, g_{22}, g_{12}$
these FPs  are placed at the corners of
a cube deformed in the $g_{12}$ direction  (see figure \ref{fig4}).
Their $\varepsilon$-expansions read \cite{Schaefer91}:
\begin{eqnarray} g_G^* & = &
\frac {3 \varepsilon}{2} -  \Big( J  + \frac {3}{2} \zeta(3)\Big)
 \frac{3\varepsilon^3}{8}, \label{4.8} \\
g_U^* & = & \frac {9 \varepsilon}{8} + \frac {39 \varepsilon^2}{256} +
 \Big( \frac{267}{4096} - \frac {693}{1024} \zeta(3) -
 \frac{189}{512} J \Big) \varepsilon^3, \label{4.9} \\
g_S^* & = & \frac {3 \varepsilon}{4} + \frac {15 \varepsilon^2}{128} +
 \Big( \frac{111}{2048} - \frac {99}{256} \zeta(3) -
 \frac{33}{128} J \Big) \varepsilon^3. \label{4.10}
\end{eqnarray}

For the evaluation of the fixed points
of the $\beta$-functions, calculated in the massive scheme
(\ref{4.3}), (\ref{4.4}) (as well as of the other quantities
of the theory) one has several alternatives.
The first possibility is to introduce $\varepsilon$-expansions for the
loop integrals. For massive renormalization these are known  for the
one- and two-loop integrals (see \cite{Amit84}):
\begin{equation}
D_2^m = \frac{1}{\varepsilon}\Big(1- \frac{\varepsilon}{2}\Big) +
O(\varepsilon), \hspace{0.5em}
  i_1 = \frac{1}{2} + \frac{\varepsilon}{4} + O(\varepsilon^2),
  \hspace{0.5em}
i_2 =  -\frac{\varepsilon}{8} + O(\varepsilon^2).
\label{4.11}
\end{equation}

A second possibility is to proceed directly in fixed dimension of
space $d=2$ or $d=3$ substituting numerical values for the
loop integrals \cite{Nickel77,Holovatch94}. For the massive  two
loop $\varepsilon$-expansion the fixed points read:
\begin{eqnarray}
g_{\rm S}^*&=& \frac {3 \varepsilon}{4} + \frac {111
 \varepsilon^2}{128}, \\
 g_{\rm U}^*&=& \frac {9 \varepsilon}{8} +
 \frac {327 \varepsilon^2}{256}, \\
g_{\rm G}^* &=& \frac {3 \varepsilon}{2} +
\frac {3 \varepsilon^2}{2}.
\end{eqnarray}
Only the first order of these results coincides with the fixed point
values of the massless renormalization scheme \cite{Schaefer91} It is
well known that the values of $\beta$-functions, fixed points, and
other intermediate functions in general depend on the RG scheme, only
the critical exponents and other observables will be independent of
the scheme followed.

We will now study the expressions (\ref{4.3}),
(\ref{4.4}) directly at fixed dimension. In this scheme the usual
way of finding the fixed points of $\beta$-functions of models
with several couplings involves the
numerical solution of the system of equations (\ref{4.5}).
To this end the asymptotic series in the coupling constants are
represented in the form of corresponding resummed expressions
$\beta^{res}$ \cite{note4}.  However, the numerical solution of the
resummed fixed point equation in general leads to inconsistent
results, as we will show in Sec. \ref{VI}.  An alternative to this
procedure and thus a third possibility to proceed was
originally proposed by Nickel and may be called a pseudo-$\varepsilon$
expansion\cite{Nickeleps}. As to our knowledge it has until now not
been applied to theories with several couplings (see
\cite{LeGuillou80}), although it seems a convenient tool to circumvent
the specific difficulties arising for the massive approach.
To apply this method,  we introduce the ``pseudo-epsilon''
parameter $\tau$ into the expressions for the $\beta$-functions
$\beta^m_{v_{aa}}, \beta^m_{v_{12}}$ in (\ref{4.3}), (\ref{4.4}) in
the following way:
\begin{eqnarray}
&&-\beta^m_{v_{aa}}/((4-d) v_{aa})
=  \tau -{\frac{4\,v_{aa}}{3}} + \ldots, \hspace{2em} a=1,2, \nonumber
\\ &&-\beta^m_{v_{12}}/((4-d) v_{12}) =  \tau -\frac{1}{3} \Big
(v_{11} + v_{22} + 2 v_{12} \Big ) + \ldots \hspace{1em}.
\label{4.12}
\end{eqnarray}
We solve for the fixed point solutions as series in
$\tau$.  The resulting series for the fixed points then either can be
resummed (to obtain the numerical values of the fixed points)
or they can be substituted into the expansions for the observables of
the theory. In the final results we substitute $\tau=1$.

Performing this procedure we get the fixed point values as series in
the pseudo-epsilon parameter $\tau$ up to the order $\tau^3$:
\begin{eqnarray}
\nonumber
&&v_G =
3/2 \tau +
  (3\, i_{1}-3/2   )\tau^2 +
  (-{\frac {9\,{i_{6}}}{8}}-
{\frac {9\, i_{1}}{4}}+
3/8+
12\, { i_{1}}^{2}-
{\frac {9\,{i_{4}}}{2}}-
\\&&
{\frac {27\,{i_{5}}}{8}}-
{\frac {9\,{i_{7}}}{8}}  ){\tau}^{3},
\label{4.13} \hspace{20em}
\\
&&v_U =
{\frac {9\tau}{8}} +
  ({\frac {93\, i_{1}}{32}}-
{\frac {93}{64}}+
{\frac {3\, i_{2}}{32}}  )\tau^2 -
  ({\frac {387\,{i_{6}}}{512}}+
{\frac {9\,{i_{8}}}{64}}+
{\frac {33\, i_{2}}{64}}+
{\frac {1281\, i_{1}}{256}}-
\nonumber \\ &&
{\frac {459}{512}}+
{\frac {693\,{i_{7}}}{512}}+
{\frac {27\, i_{3}}{128}}+
{\frac {369\,{i_{4}}}{64}}+
{\frac {1485\,{i_{5}}}{512}}-
{\frac {3\,{ i_{2}}^{2}}{64}}-
{\frac {9\, i_{2}\, i_{1}}{8}}-
\nonumber \\ &&
{\frac {969\,i_{1}^2}{64}} -
{\frac {27\,d i_{2}}{512}}  ){\tau}^{3},
\hspace{20em}
\label{4.14}
\\
&& v_S =
3/4 \tau+
  ({\frac {3\, i_{2}}{16}}+
{\frac {33\, i_{1}}{16}}-
{\frac {33}{32}}  )\tau^2 +
  (3/4-
{\frac {27\, i_{2}}{32}}-
{\frac {261\,{i_{5}}}{128}}-
\nonumber \\ &&
{\frac {135\,{i_{4}}}{32}}-
{\frac {261\, i_{1}}{64}}-
{\frac {63\,{i_{6}}}{128}}+
{\frac {3\,{ i_{2}}^{2}}{32}}+
{\frac {33\, i_{2}\, i_{1}}{16}}+
{\frac {9\,d i_{2}}{128}}-
{\frac {99\,{i_{7}}}{128}}-
{\frac {9\,{i_{8}}}{32}}-
\nonumber \\ &&
{\frac {9\, i_{3}}{32}}+
{\frac {363\,{ i_{1}}^{2}}{32}}  ){\tau}^{3}.
\hspace{20em}
\label{4.15}
\end{eqnarray}

The expressions (\ref{4.8}) - (\ref{4.10}) \cite{Schaefer91} and
the expressions (\ref{4.3}) - (\ref{4.15}) give
the fixed point values of ternary solutions in the  massless and
massive renormalization schemes and are the main results to be used in
the subsequent calculations.

Looking for the stability of the above described fixed points one
finds that only the fixed point {\it S} is stable \cite{Schaefer91}.
In the excluded
volume limit of infinitely long chains the behavior of a system of two
polymer species is thus described by the same scaling laws as a
solution of only one polymer species.
Nevertheless taking into account that
real polymer chains are not infinitely long one may also find
crossover phenomena which are governed by the unstable
fixed points. Knowing the complete RG flow allows to
describe crossover phenomena in the whole accessible region
\cite{Schaefer91}.
However, for the purpose of our study we are interested only in the
values of the fixed points and the properties of the star vertex
functions at these fixed points.

\section{Results for Exponents}
\label{V}
For homogeneous stars of polymer chains of one species alone, several
sets of star exponents have been defined, each describing either the scaling
properties of the configurational number (see formula (\ref{1.3})
of this article), or the anomalous dimensions of star vertices, etc. Due to
scaling relations, these exponents can be expressed in terms
of each other \cite{Duplantier89}. In this sense, each set of star
exponents forms a complete basis. For the copolymer and MAW stars,
we here choose to present our results in terms of the
exponents
$\eta_{f_1f_2}$ and $\eta^{MAW}_f$ given by the fixed point values of
the functions $\eta_{*f_1f_2}(g_{ab})$ (\ref{3.14}) and
$\eta^{{\rm MAW}}_f(g_{ab})$ (\ref{3.15}).
Let us define the asymptotic values of
copolymer star exponents and MAW star exponents by:
\begin{eqnarray} \label{5.1}
\eta^S_{f_1f_2} &=& \eta_{*f_1f_2}(g_{ab})|_S ,
\\ \label{5.2}
\eta^G_{f_1f_2} &=& \eta_{*f_1f_2}(g_{ab})|_G ,
\\ \label{5.3}
\eta^U_{f_1f_2} &=&
\eta_{*f_1f_2}(g_{ab})|_U=
\eta_{*f_2f_1}(g_{ab})|_{U^{\prime}},
\\ \label{5.4}
\eta^{{\rm MAW}}_f &=& \eta^{{\rm MAW}}_f(g_{ab})|_G.
\end{eqnarray}
The exponent in the symmetrical FP S can also be expressed by
$\eta^S_{f_1f_2}=\eta^U_{f_1+f_2,0}$.
Starting from the expressions for the fixed points given in the
previous section,
and the relations (\ref{5.1}) - (\ref{5.4}), we find the series
for the  star exponents.
In $\epsilon$-expansion we obtain the following
expansions for $\eta_{f_1f_2}$:
\begin{eqnarray}
&&\eta^G_{f_1f_2} (\varepsilon) =
- f_1 \, f_2 \frac {\varepsilon}{2}+
f_1 \, f_2 \,\Big ( f_2 - 3 + f_1 \Big )
\frac{\varepsilon^2}{8}-
f_1 \, f_2 \Big ( f_2 - 3 + f_1 \Big )
\Big ( f_1 +
\nonumber \\ &&
f_2 + 3 \, \zeta (3)-
3 \Big )\frac {\varepsilon^3}{16},
\label{5.8}
\end{eqnarray}
\begin{eqnarray}
&&\eta^U_{f_1f_2}(\varepsilon) =
{\it f_1}\Big (1-{\it f_1}-
3{\it f_2}\Big )\frac {\varepsilon}{8} +
{\it f_1}\Big (25-33{\it f_1}+
8{{\it f_1}}^{2}-
91{\it f_2}+
42{\it f_1}{\it f_2}+
\nonumber \\ &&
18{{\it f_2}}^{2}\Big) \frac {\varepsilon^2}{256}+
{\it f_1}\Big (577-
969{\it f_1}+
456{{\it f_1}}^{2}-
64{{\it f_1}}^{3}-2463{\it f_2}+
2290{\it f_1}{\it f_2}-
\nonumber \\ &&
492{{\it f_1}}^{2}{\it f_2}+
1050{{\it f_2}}^{2}-
504{\it f_1}{{\it f_2}}^{2}-
108{{\it f_2}}^{3}-
712\zeta (3)+
936{\it f_1}\zeta (3) -
\nonumber \\ &&
224{{\it f_1}}^{2}\zeta (3)+
2652{\it f_2}\zeta (3)-
1188{\it f_1}{\it f_2}\zeta (3)-
540{{\it f_2}}^{2}\zeta (3)\Big )
\frac {\varepsilon^3}{4096},
\label{5.9}
\end{eqnarray}
\begin{eqnarray} &&
\eta^{MAW}_{f_1f_2}(\varepsilon) =
- ({\it f_1}-
1 ){\it f_1} \frac {\varepsilon}{4}+
{\it f_1} ({\it f_1}-
1 ) (2{\it f_1}-
5 ) \frac {\varepsilon^2}{16}-
 ({\it f_1}-
1 ){\it f_1} (4{{\it f_1}}^{2}-
\nonumber \\
&& 20{\it f_1}+
8{\it f_1}\zeta (3)-
19\zeta (3)+
25 )
\frac {\varepsilon^3}{32}.
\label{5.10}
\end{eqnarray}
Here $\zeta(3)\simeq 1.202$ is the Riemann $\zeta$-function.
The above formulas reproduce the 3rd order calculations of
the scaling exponents of homogeneous polymer stars
$\gamma_f-1=\nu(\eta^{\rm U}_{f,0}-f\eta^{\rm U}_{2,0})$
\cite{Schaefer92}. The exponents $\lambda^{\rm (xx)}$
given to 2nd order in equations ${\rm (xx)}$ of
\cite{Cates87} to describe the multifractal scaling properties of a
Laplacian field with fractal boundary conditions are reproduced
following $\lambda^{(29)}(n)=-\eta^{\rm G}_{2,n}$,
$\lambda^{(47)}(n)=-\eta^{\rm U}_{2,n}+\eta^{\rm U}_{2,0}$,
$\lambda^{(48)}_{\rm e}(n)=-\eta^{\rm G}_{1,n}$,
$\lambda^{(49)}_{\rm e}(n)=-\eta^{\rm U}_{1,n}$,
correcting a misprint in eq.(49) of \cite{Cates87}.
Also the 2nd order results for exponents
$x_{L,n}-x_{L,1}=-2(\eta^{\rm G}_{L,n} - \eta^{\rm G}_{L,1}) $
of \cite{Duplantier91} and the MAW exponents
$\sigma_L=1/2\eta^{\rm MAW}_L $ defined in \cite{Duplantier88}
find their 3rd order extension by the above expansions.
The pseudo-$\varepsilon$ expansions for $\eta_{f_1f_2}$ obtained in
the massive scheme read:
\begin{eqnarray}
\nonumber &&
\eta^G_{f_1f_2}= \tau\,\epsilon
\frac{f_1f_2}{2} \,\Big \{-1+
(f_{{2}}-3+f _{{1}} )\Big [\tau\, (i_{{1}}-1/2)+
\frac { \tau^2} {8}\Big (32\, (i_{{1}}-1/2 )^{2}+6\,i_{{6}}-18\,
i_{{4}} - 8 +
\\ \nonumber
&& 22\,i_{{1}} - 6\,i_{{7}}-6\,i_{{5}}- (f_2-3+f_1)
(2+6\,i_{{4}}-6\,i_{{1}}) \Big ) \Big ] -
{\frac {\tau^2}{4} (f_{{1}}f_{{2}}-2 ) (
1+3\,i_{{5}}+3\,i_{{6}}-6\,i_{{1}} )}\Big \} ,
\\ \label{5.11}
\end{eqnarray}
\begin{eqnarray}
\nonumber &&
\eta^U_{f_1f_2}=
\tau\,\epsilon\,\frac{f_1}{1024}
\{
128-128\,f_{{1}}-384\,f_{{2}}
+\tau[
\left (288\,i_{{1}}-144\right ){f_{{2}}}^{2}-208+416\,i_{{1}}+32\,
i_{{2}}+
\\ \nonumber
&&
\left (272-32\,i_{{2}}-544\,i_{{1}}\right )f_{{1}}+\left (
-32\,i_{{2}}-1472\,i_{{1}}+736\right )f_{{2}}+\left (-64+128\,i_{{
1}}\right ){f_{{1}}}^{2}+
\\ &&
\left (-336+672\,i_{{1}}\right )f_{{1}}f_
{{2}} ]
+\tau^2\sum_{k_1k_2}f_1^{k_1}f_2^{k_2}\eta_{U;k_1;k_2}
\},
\label{5.12}
\end{eqnarray}
\begin{eqnarray} \nonumber
&&
\eta^{MAW}_f= \tau\,\epsilon\,\frac{f(f-1)} {16}
\Big \{-4 - (20\,i_{{1}}-10+4\,f-8\,fi_{{1}})\tau+
\Big [(-3\,i_{{5}}+18\,i_{{1}}-12\,i_{{4}}-3\,i_{{6}}-
\\ \nonumber
&&
5){f}^{2}+ (3\,i_{{5}}+32\,{i_{{1}}}^{2}-8\,i_{{7}}+15
\,i_{{6}}-88\,i_{{1}}+42\,i_{{4}}+23)f-26-18\,i_{{6}}+19\,i_{{7}}+
12\,i_{{5}}+
\\ &&
106\,i_{{1}}-80\,{i_{{1}}}^{2}-30\,i_{{4}}
\Big ]{\tau}^{2} \Big \}.
\label{5.13}
\end{eqnarray}
The expressions for the three loop terms $\eta_{U;k_1;k_2}$ in
(\ref{5.12}) are given in the Appendix B.
It has been pointed out in \cite{Cates87} that for the exponent
$\eta_{12}^G=-\lambda^{(29)}(1)$ (see above) an exact estimate equal
to our first order contribution may be found. It is indeed
remarkable that all higher order contributions to $\eta_{12}^G$ appear
to vanish in both approaches.

With these exponents we can describe the scaling behavior of
polymer stars and networks of two components, generalizing
the relation for single component networks \cite{Duplantier89}.
In the notation of (\ref{1.3a})
we find for the number of configurations of a network $\cal G$
of $F_1$ and $F_2$ chains of species $1$ and $2$
\begin{equation}
\label{20a}
{\cal Z}_{\cal G} \sim
(R/\ell)^{\eta_{\cal G} -F_1\eta_{20}-F_2\eta_{02}} \mbox{, \, \,with
} \eta_{\cal G} = -d L + \sum_{f_1+f_2\geq 1} N_{f_1f_2}\eta_{f_1f_2},
\end{equation}
where $L$ is the number of Loops and $N_{f_1f_2}$ the number of
vertices with $f_1$ and $f_2$ arms of species $1$ and $2$ in
the network $\cal G$. To receive an appropriate scaling law
we assume the network to be built of chains which for both species
will have a coil radius $R$ when isolated.

For the sake of completeness we give also the results in
$\varepsilon$- and pseudo-$\varepsilon$ expansions for the correlation
length critical exponent $\nu = 1/(2 + \eta_{20} - \eta_{\phi})$ and
the pair correlation function critical exponent $\eta_\phi$ in the non
trivial fixed point:
\begin{eqnarray}
\label{5.14} &&
\eta_{\phi}(\varepsilon) =
{\frac {{\varepsilon}^{2}}{64}}+{\frac {17{\varepsilon}^{3}}{1024}},
\\ \label{5.15}&&
\nu(\varepsilon) = {\frac {1}{2}}+{\frac {1}{16}}\varepsilon+
{\frac {15}{512}}{\varepsilon }^{2}+\left ({\frac {135}{8192}}-
{\frac {33\zeta (3)}{1024}} \right ){\varepsilon}^{3},
\\ \label{5.16} &&
\eta_{\phi}(\tau) = \frac {-(4-d)}{128}\tau
 (16\tau{\it i_{2}}-12{\tau}^{2}{\it i_{2}}-24{
\tau}^{2}{\it i_{8}}+
 8{\tau}^{2}{{\it i_{2}}}^{2}+
88{\tau}^{2}{\it i_{1}}
{\it i_{2}} ), \hspace{1em}
\\ \label{5.17} &&
\nu(\tau) =  \frac {1}{2}+\frac {4-d}{16} \tau-
\frac{(4-d)}{512}(4-40 i_{1}+8 i_{2} + 4d) \tau^2 -
\nonumber \\  &&
\frac{(4-d)}{512}
 (6+12 i_2+10 i_1+84 i_4+63 i_5+3 i_6+
33 i_7-12 i_8-
\nonumber \\  &&
220 i_1^2+ 4 i_2^2+24 i_1 i_2 - d+10 i_1 d-
 2 i_2d -\frac {d^2}{2}  ) \tau^3.
\end{eqnarray}

\section{Resummation}\label{VI}
As is well known, the perturbation series expansions of renormalized
field theory are non-convergent but generally assumed to be
asymptotic. For the exponents $\eta_{f_1f_2}$ this behavior is
indicated in the corresponding columns of the tables \ref{tab3} and
\ref{tab4}, where the series are summed without further analysis.

The increase in the coefficients of the high order terms of
perturbation theory series may
be estimated using information such as the combinatorial growth of the
contributions with order. The series for the
$\beta$ function of the $O(m)$ symmetric $\phi^4$ model with one coupling
$g$ has the following asymptotic behavior \cite{Brezin77,Lipatov77}:
\begin{eqnarray}
\beta(g) &=& \sum_k A_k g^k, \label{6.1}\\
A_k &=& c k^{b_0} (-a)^k k! [1 + O(1/k)] \hspace{1em},
k\to\infty. \label{6.2}
\end{eqnarray}
The quantities $a,b_0,c$ were calculated in \cite{Brezin77,Brezin78}.
A similar behavior is found for the critical exponents
expressed as a series in powers of the coupling. The same results also
apply to the divergence of the $\varepsilon$ and pseudo
$\varepsilon$ expansions derived above. The property (\ref{6.2})
indicates the Borel summability of the series $\beta(g)$
\cite{Hardy48}. The Borel resummation procedure
takes into account the asymptotic behavior of the coefficients and
maps the asymptotic series to a convergent series with the
same asymptotic limit.  The function $\beta_{aa}$ (\ref{3.12})
coincides with the $O(m)$ symmetric $\beta$ function
(\ref{6.1}) in the polymer limit $m=0$. So its asymptotic behavior is
known.  The asymptotic behavior of the off diagonal
$\beta$ function $\beta_{12}$ was found by instanton analysis (see
\cite{ZinnJustin89,Itzykson80}) in \cite{Schaefer91}.

Let us introduce the techniques for resummation, using the known
asymptotic behavior of the series.  Here, we apply Pad\'e-Borel
resummation \cite{Nickel78} and a resummation refined by a
conformal mapping \cite{Zinn81}. The first way of resummation is
applicable only for alternating series, while the second one is more
universal.  The resummation procedures are as follows
\cite{Nickel78,LeGuillou80,Hardy48,Zinn81}. For an asymptotic
series
\begin{equation} \label{6.3}
f(\varepsilon)= \sum_{j}
f^{(j)}\varepsilon^{j},
\end{equation}
one defines the Borel-Leroy transform $f^B(\varepsilon)$ of the series
by:
\begin{equation} \label{6.4}
f^B(\varepsilon)= \sum_{j} \frac{f^{(j)} \varepsilon^{j}}
{\Gamma(j+b+1)},
\end{equation}
with the Euler $\Gamma$-function ($b$ is a fit parameter).
Then, the initial series may be regained
from
\begin{equation} \label{6.5}
f^{res}(\varepsilon)= \int_{0}^{\infty}d t t^b e^{-t}
f^B(\varepsilon t).
\end{equation}
Substituting for $f^B(\varepsilon)$ its analytic continuation in form
of a Pad\'e approximant and evaluating (\ref{6.5}) for the
truncated series,
this procedure constitutes the Pad\'e Borel resummation
\cite{Nickel78,Hardy48}.
The conformal mapping technique in addition uses
the constant $a$ in (\ref{6.2}). Assuming the
behavior (\ref{6.2}) holds also for the expansion of
$f(\varepsilon)$ in $\varepsilon$,  one concludes that the
singularity of the transformed series $f^B(\varepsilon)$ closest
to the origin is located at the point $(-1/a)$.
Conformally mapping the $\varepsilon$ plane onto a disk of radius 1
while leaving the origin invariant,
$$ w =
\frac{(1+a\varepsilon)^{1/2}-1}{(1+a\varepsilon)^{1/2}+1},
\hspace{3em}
\varepsilon= \frac{4}{a} \frac{w}{(1-w)^2},
$$
substituting this into $f^B(\varepsilon)$ and expanding in $w$
we receive a series defined on the disk with radius $1$ on the $w$
plane. This series is then resubstituted into  (\ref{6.5}). In order
to weaken a possible singularity in the $w$-plane the corresponding
expression is multiplied by  $(1-w)^{\alpha}$ introducing an
additional parameter $\alpha$ \cite{Zinn81}. In the resummation
procedure the value of $a$ is taken from the known high-order
behavior of the $\varepsilon$-expansion series while $\alpha$ is
chosen in our calculations as a fit parameter defined by the condition
of minimal difference between resummed 2nd order and 3rd order
results. The resummation procedure was seen to be quite insensitive to
the parameter $b$ introduced in the Borel-Leroy transformation
(\ref{6.4}) \cite{LeGuillou80}.

For the resummation of the exponents $\eta_{f_1f_2}$ we take into account
the combinatorial factors which multiply each
contribution according to the numbers of chains $f_1$ and $f_2$. We
include  an additional factor $(f_1+f_2)^k$ for the $k$th order
contributions, multiplying the constant $a$ by $(f_1+f_2)$.  For
resummation of the series at the fixed points S,G and U the following
values of $a=a^{\rm S},a^{\rm G},a^{\rm U}$ are used
\cite{Brezin77,Schaefer91}:
\begin{equation}
a^{\rm S} = a^{\rm G} = 3/8  \hspace{1em}
\mbox { and } \hspace{1em}
a^{\rm U}= 27/64.
\end{equation}
By analogy we use the same procedures developed for the $\varepsilon$
expansion also for the $\tau$-expansion which we assume to have the
same asymptotic behavior as it is in the same way collecting
contributions of the same loop order.

Let us note here that the conventional resummation of the $\beta$-
functions in the massive approach leads to a severe inconsistency which is
the reason for us to take the pseudo-$\varepsilon$ or
$\tau$-expansion method.
The distinct
feature of the $\beta_{ab}$-functions introduced here
is that they are functions of different numbers of variables which
leads to ambiguities in their analytical continuation via Pad\'e
approximants or rational approximants of several variables (see
\cite{Baker81}). Let us illustrate this for the example of
the two-loop approximation. The corresponding expressions read:
\begin{eqnarray}
\label{6.7}
\beta_{v_{aa}} &=& -(4-d) v_{aa} f_{v_{aa}}(v_{aa}),
\hspace{1em} a=1,\,2,
\\  \label{6.8}
\beta_{v_{12}} &=& -(4-d) v_{12}
f_{v_{12}}(v_{11}, v_{22}, v_{12}),
\end{eqnarray}
with obvious expressions for $f_{v_{11}},f_{v_{22}},f_{v_{12}}$.
In order to obtain the analytical continuation of the Borel
transformed functions  of one variable $f_{v_{aa}}(v_{aa})$
(\ref{6.7}) one can make use of the $[1/1]$ Pad\'e approximant.
Solving the corresponding non linear equations numerically we find for
the non-trivial fixed point $S$: $v_{11}=v_{22}=1.1857$ \cite{note5}.
In order to apply a similar resummation technique to the function
$f_{v_{12}}(v_{11}, v_{22}, v_{12})$
(\ref{6.8}) one can make use of a generalization of Pad\'e
approximants to the case of several variables, i.e. represent the
Borel transform $f^B_{v_{12}}$ of $f_{v_{12}}$
in the form of a rational approximant $f^P$ of three variables
\cite{Baker81}:
\begin{equation}
f^P_{v_{12}}(v_{11}t, v_{22}t, v_{12}t) =
\frac{1+ a_1(v_{11}, v_{22}, v_{12}) t  +
a_2(v_{11}, v_{22}, v_{12}) t^2} {1+ b(v_{11}, v_{22}, v_{12})t}.
\label{6.9}
\end{equation}
In spite of the fact that the rational approximant (\ref{6.9})
preserves the projection properties of the initial series (\ref{6.8}),
i.e.  putting any pair of variables $\{v_{11}, v_{22}, v_{12}\}$
to zero in (\ref{6.9}) one gets the appropriate $[1/1]$ Pad\'e
approximant  for the remaining variable, the
``global'' symmetry is not preserved. Due to different
analytical continuations for the Borel transforms of the series
(\ref{6.7}) on one hand and of (\ref{6.8}) on the
other, solving the fixed-point equation for the resummed function
\begin{equation}
f_{v_{12}}^{res}=0
\label{6.10}
\end{equation}
we will never obtain a symmetrical solution
$v_{11}^*=v_{22}^*=v_{12}^*\neq 0$.
For the fixed point $S$ we substitute
$v_{11}=v_{22}=1.1857$ and solving (\ref{6.10})
we receive $v_{12}=0.9571$ (!) \cite{note6}.
The reason is, that substituting {\it numerical} values of fixed point
coordinates $v_{11},v_{22}$ into  (\ref{6.10}) we lose information
about the contributions to the fixed point value from different
orders of the perturbation theory series.
So it appears quite natural to restore this information by
generalizing the pseudo-$\varepsilon$ expansion \cite{Nickeleps}
to the case
of several couplings as described in section \ref{IV}.

\section{Numerical Results}
\label{VII}

In the following we present our numerical results for the exponents
$\eta_{f_1f_2}^{\rm G}$, $\eta_{f_1f_2}^{\rm U}$ and $\eta^{MAW}_f$.
The exponent in the symmetrical fixed point $S$ is included due to the
relation  $\eta_{f_1f_2}^{\rm S} = \eta_{f_1+f_2,0}^{\rm U}$
\cite{note7}. Numerical results for the exponent
$\gamma_f-1=\nu(\eta_{f0}^{\rm U}-f\eta_{20}^{\rm U})$ may be found in
$\varepsilon$-expansion in  \cite{Schaefer92} and in
pseudo-$\varepsilon$ expansion in \cite{FerHol96a}.

\subsection{$d=3$}

Let us first consider the case $d=3$. The tables \ref{tab3} and
\ref{tab4} show some of the resummed results for the $\varepsilon$-
and $\tau$ expansions in comparison with the naive resummation of the
series. While the non-resummed results differ to a great extent for 
the two approaches at high $f_1,f_2$, resummation shows that the two
schemes yield consistent numerical estimates. Tables \ref{tab5},
\ref{tab6}, and \ref{tab7} list our final results using the
resummation procedure refined by the conformal mapping technique as
described in the previous section.

Comparing the numerical values listed in the above tables it is
convincing that the two approaches and the different resummation
procedures all lead to results which lie within a bandwidth of
consistency, which is broadening for larger values of number of chains
$f_1,f_2 >1$.
This is not surprising as we have seen in  section \ref{V} that
our expansion parameters are multiplied by $f_1$ and $f_2$.
Rather it is remarkable that even for a total number of chains of the
order of 10 (see tables \ref{tab5}, \ref{tab6}) we still receive results
which are comparable to each other.

It seems noteworthy that at least for low numbers of chains
($f_1+f_2\sim 4$) the non resummed $\tau$-expansion seems to give
results which do not differ essentially from the resummed values.
Also the non refined Pad\'e-Borel results of the $\tau$-expansion are
closer to the refined summation of the $\varepsilon$-expansion.

\begin{figure} [H]
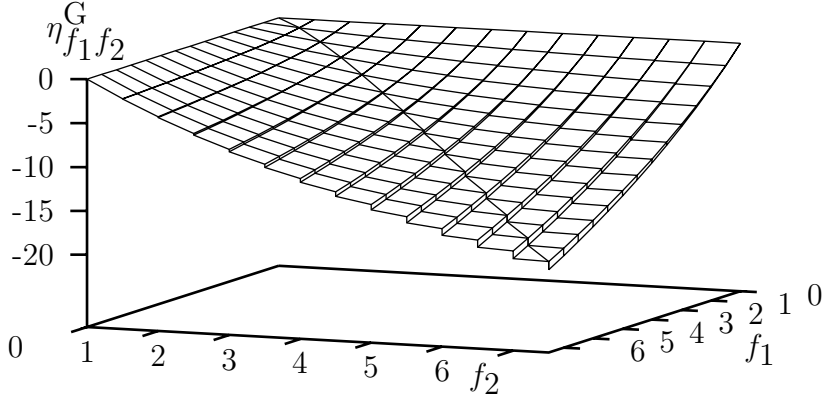

\input fig5.pic
\caption{Exponent $\eta^G_{f_1f_2}$ in the `Gaussian' fixed point
at $d=2$ obtained in $\varepsilon$-expansion and in fixed $d$ scheme.
Steps on the ``flying carpet'' correspond to the difference
of the results of the two renormalization group approaches.
The line across traces the diagonal values $\eta^G_{ff}$.
}
\label{fig5}
\end{figure}

Does the data answer the question of convexity of the spectrum?
A close study of the matrix of values reveals, that for fixed $f_1$
both $\eta^{\rm G}_{f_1f_2}$ and $\eta^{\rm U}_{f_1f_2} $
are convex from above as function of $f_2$, thus yielding `MF
statistics'. The relation to a MF spectral function for $f_1=1,2$ has
been pointed out in \cite{Cates87}, it is analyzed in close detail in
view of the new data and FT formulation in a separate publication
\cite{FerHol96d}.
On the other hand also copolymer stars should repel each other.
This is found to be true as well, the corresponding convexity from
below shows up e.g. along the diagonal values $\eta_{ff}$ as
function of $f$. The general relation
$\eta_{f_1f_2}+\eta_{f'_1f'_2} \geq \eta_{f_1+f'_1,f_2+f'_2} $
is always fulfilled.
Thus, even though simple power $k$ of field operators $\phi^k$ do not
describe MF moments \cite{Duplantier91},
they may be written as a power $L+k$ of field operators of suitable
symmetry which have the appropriate short distance behavior. This is
also illustrated in the next subsection by Fig. \ref{fig5}, showing
the spectrum of exponents $\eta^{\rm G}_{f_1f_2}$ in the 2D limit. The
opposite convexity along the diagonal as opposed to each of the two
axes is clearly seen for these combinations of two random walk stars
which mutually interact.

\subsection{$d=2$}

While two dimensional star polymers up to now have not found an
experimental realization, their study is of some theoretical interest.
It has been shown that the scaling dimensions of two-dimensional
uniform polymer stars belong to a limiting case of the so-called
conformal Kac table \cite{Kac79,Belavin84,Friedan84}.  They have been
calculated exactly by Coulomb gas techniques
\cite{Duplantier86,Saleur86}.  An exact relation has also
been proposed for stars of mutually avoiding walks
\cite{Duplantier88,Duplantier88a}. But it
is still an open question if exact results for the copolymer star
system may be derived in this formalism.
Our numerical results for the exponents $\eta_{f_1f_2}^{\rm G}$,
$\eta_{f_1f_2}^{\rm U}$, and $\eta_{f_1f_2}^{\rm MAW}$ are presented
in tables \ref{tab8}, \ref{tab9}, and \ref{tab7}.

Exact results for exponents of two dimensional systems which are described
by a conformal field theory with central charge $c<1$ may be taken from
the Kac table of scaling dimensions \cite{Kac79,Belavin84,Friedan84}:
\begin{equation}\label{7.1}
h_{p,q}(m)=\frac{[(m+1)p - mq]^2 - 1}{4m(m+1)},
\end{equation}
where $p,q$ are integers in the minimal block
\begin{equation}\label{7.2}
1 \leq p \leq m-1, \hspace{1em} 1 \leq q \leq p,
\end{equation}
and $m$ is connected with the central charge $c$ by
\begin{equation}\label{7.3}
c = 1 - 6/m(m+1), \hspace{1em} m \geq 3 .
\end{equation}
The exact result for the star exponents of uniform stars in two dimensions
is received in the sub-limiting case of $m=2$ (which means $c=0$) for
half integer values of $p$ \cite{Duplantier86,Saleur86} :
\begin{equation}\label{7.4}
x_f = 2h_{f/2,0} = (9 f^2 - 4)/48.
\end{equation}
The scaling dimension $x_f$ is related to the exponent
$\eta_f$ by:
\begin{equation}\label{7.5}
x_f = \frac{1}{2} f (d-2+\eta) - \eta_f.
\end{equation}
For the exponents of the star of MAW the following result was
conjectured for $d=2$ \cite{Duplantier88,Duplantier88a,Saleur92}:
\begin{equation}\label{7.6}
  \eta_f^{\rm MAW} = x_f^{\rm MAW} = 2 h_{0,f} = \frac{1-4f^2}{12}.
\end{equation}
These values are shown in the last column of table \ref{tab7}.
Plotting the resummed data for $\eta^{\rm MAW}_f$ from table \ref{tab7}
with respect to $f^2$ one finds good agreement with the conjectured
slope of $-1/3$.

The qualitative behavior of the exponent $\eta_{f_1f_2}^{\rm G}$ in the
Gaussian fixed point is shown in Fig. \ref{fig5}. The steps in the
`flying carpet' correspond to the difference of the results of the two
RG approaches. Note that the curvature of the surface along the diagonal
in the $f_1,f_2$ plane has opposite sign to that along each of the axes.
From this curvature it is obvious that the dependence of the exponent
on $f_1,f_2$ may not be described by a simple second order polynomial.
The best fit we could find to our resummed data using a simple formula which
reproduces the vanishing result for $f_1+f_2 = 3$ found in
$\varepsilon$-expansion reads:
\begin{equation}\label{7.7}
\eta_{f_1f_2}^{G,app} = - f_1f_2(a+b/(f_1+f_2)),\hspace{1em}
a = 1/4, \hspace{1em} b= 3(1-a).
\end{equation}
Note that the right hand side of (\ref{7.7}) vanishes if $f_1$ or
$f_2$ is zero according to our perturbative results. This might be a
defect of the perturbation theory as a finite result may be expected
in $d=2$ as in equations (\ref{7.4}),(\ref{7.6}) evaluated for $f=0$.

In 2D however, each chain of a star will interact only with its direct
neighbors. A star described here by $\eta^{\rm G}_{ff} $ will behave
like a MAW $2f$-star if each species-1 chain has two neighbors of
species-2 whereas it will behave differently if the chains are ordered
such that each species is in one bulk of chains. The 2D
copolymer stars in this sense reveal an even richer behavior.  Thus,
the copolymer generalization of the MAW star adds another problem, for
which a rigorous formulation in terms of an exactly solvable 2D model
is yet to be found.



\section{Conclusions and Outlook}\label{VIII}
Several reasons motivated our study. First, we intended to reveal the
scaling behavior of copolymer stars and networks in solutions
generalizing former studies of homogeneous polymer networks. This
included revisiting the theory of ternary polymer solutions and
adding an
independent approach to the calculations. Secondly, the description of
multifractal spectra in terms of random walks \cite{Cates87} promised to
prove the relation of field theory and multifractals for this case.
In particular we intended to check the convexity properties expected
for the spectrum of exponents. A third motivation arose from the known
peculiarities of polymers and polymer stars in two dimensions. Apart
from numerically verifying previous results on polymer and mutually
avoiding walk stars we pose the question of finding an exactly
solvable (conformal) two dimensional theory for general copolymer
stars.

We have extensively studied the spectrum of exponents governing the scaling
properties of stars of walks taking into account the self and mutual
interactions of a system of species of polymers.
Our study was performed in the framework of field theoretical RG
theory using two complementary approaches: The renormalization at zero
mass in conjunction with $\varepsilon$ expansion and massive
renormalization at fixed dimension with numerical evaluation of loop
integrals.  We have formulated the problem of finding the scaling
exponents of stars of interacting and non interacting walks in terms
of the determination of the scaling dimensions of composite field
operators of Lagrangian field theory. On the one hand this allows for
the application of well developed formalisms and methods for
analyzing the scaling properties. On the other hand this defines these
new families of exponents extending previous sets in the
framework of Lagrangian field theory.  Our results agree with the
previous studies of special cases which were in part done only to 2nd
order of the $\varepsilon$ expansion. We have here considered the
general case of a star of two mutually avoiding sets of walks, the
walks of each set either self-interacting or not. Also we have studied
the case of a star of mutually interacting walks. All calculations
were performed to third order of perturbation theory. The sets of
exponents are given in $\varepsilon$-expansion (formulas (\ref{5.8}) -
(\ref{5.10})) and in
terms of a pseudo-$\varepsilon$-expansion (\ref{5.11}) - (\ref{5.13})).
The latter has proven to be a most suitable tool to evaluate this
massive theory containing serveral couplings.  We have shown that the
conventional way of direct solution even of the resummed expressions
for the fixed points of the theory would lead to severe problems in
this case.  We have evaluated the series obtained in both approaches
for space dimensionality $d=2$ and $d=3$. Numerical values are
produced by careful resummation of the asymptotic series using the
results of an instanton analysis of the three coupling problem
\cite{Schaefer91}.  For comparison we have also given the results of
naive summation as well as standard Pad\'e Borel resummation for
selected cases.

We have found remarkable consistency and stability of the results in
$d=2$ and $d=3$ with expected growing of deviations for large number of
arms of one star. The same methods were applied previously to the
problem of uniform star polymers and have led to results
\cite{Schaefer92,FerHol96a} in good agreement with Monte Carlo (MC)
simulations \cite{Barrett87,Batoulis89}.
We hope our present calculations might also stimulate MC studies of
the copolymer star problem.

The study we performed for two dimensions might have no direct
application to the physics of real polymers but it could perhaps give
some insight to the problem of mapping our theory to a two dimensional
conformal field theory.  The resummed values of the exponents for
stars of mutually avoiding walks are in fair agreement with an exact
result previously conjectured \cite{Duplantier88,Duplantier88a}.
 The exponents for the case
of stars of two mutually avoiding sets of walks on the other hand show
a dependence on the numbers of walks which may not
be described by a second order polynomial as derived from the general
Kac formula \cite{Kac79,Belavin84,Friedan84}. This may be seen
already qualitatively from the fact that the curvature of the function
$\eta_{f_1f_2}$ (see Fig. \ref{fig5}) of the two variables $f_1,f_2$
along each of the axes in the $f_1f_2$- plane has the opposite sign to
the curvature along the diagonal $f_1=f_2$.

It is this fact on the other hand, which shows that the series of
exponents $\eta_{f_1f_2}$ is a good candidate
for finding its application in the theory of multifractal (MF) spectra
\cite{Halsey86}.  The MF spectrum describing the moments of a fractal
probability measure fulfills exact conditions of convexity.  Deriving
such a MF spectrum on the other hand from the scaling dimensions of a
series of composite field operators is only possible if the scaling
dimensions show the appropriate convexity \cite{Duplantier91}. This in
fact is given for our case and the series of exponents may be related
to the MF spectrum generated by harmonic diffusion near an absorbing
fractal \cite{Cates87}.
This also allows for a field theoretic test of results
for the short-distance correlations on multifractals
\cite{Ferber97}.
This relation and the calculation of the MF
spectrum on the basis of the results presented here, is subject of a
separate publication \cite{FerHol96d}.

\section*{Acknowledgements}
It is a pleasure to acknowledge useful discussions with Lothar Sch\"afer,
Essen. This work was supported in part by SFB 237 of Deutsche
Forschungsgemeinschaft and by the Ukrainian Foundation of Fundamental
Studies (grant No 24/173).  C.v.F. thanks for hospitality at the
School of Physics, Tel Aviv University and support by a Minerva
fellowship.

\begin{appendix}
\renewcommand{\thesection}{\Alph{section}}%
\renewcommand{\theequation}{\Alph{section}.\arabic{equation}}%
\section{ Loop Integrals. Graphs, $\varepsilon$-expansion,
Numerical Values
}\label{A}

\begin{figure} [htbp]
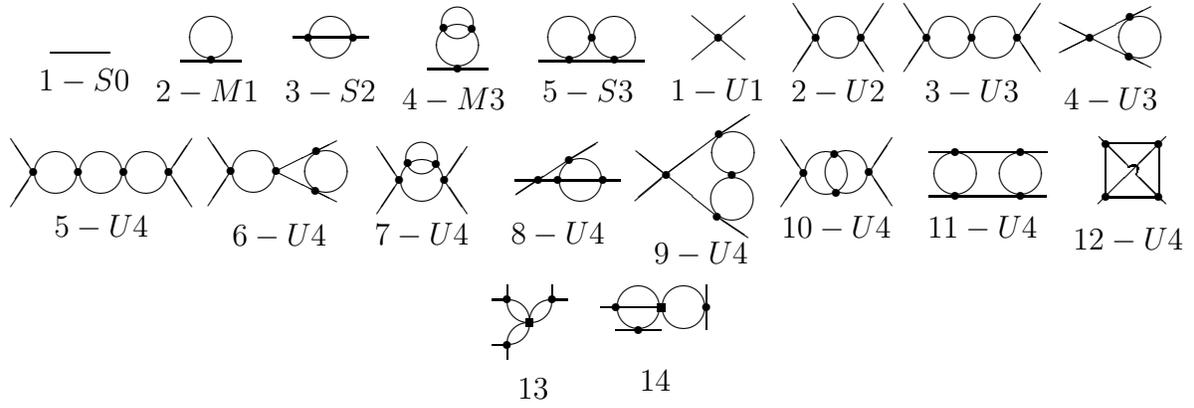

\begin{center}
\unitlength1mm
$\begin{array}{c} \input{1-S0.pic}\\ 1-S0 \end{array}$
\unitlength1mm
$\begin{array}{c} \input{2-M1.pic}\\ 2-M1 \end{array}$
\unitlength1mm
$\begin{array}{c} \input{3-S2.pic}\\ 3-S2 \end{array}$
\unitlength1mm
$\begin{array}{c} \input{4-M3.pic}\\ 4-M3 \end{array}$
\unitlength1mm
$\begin{array}{c} \input{5-S3.pic}\\ 5-S3 \end{array}$
\unitlength1mm
$\begin{array}{c} \input{1-U1.pic}\\ 1-U1 \end{array}$
\unitlength1mm
$\begin{array}{c} \input{2-U2.pic}\\ 2-U2 \end{array}$
\unitlength1mm
$\begin{array}{c} \input{3-U3.pic}\\ 3-U3 \end{array}$
\unitlength1mm
$\begin{array}{c} \input{4-U3.pic}\\ 4-U3 \end{array}$
\unitlength1mm
$\begin{array}{c} \input{5-U4.pic}\\ 5-U4 \end{array}$
\unitlength1mm
$\begin{array}{c} \input{6-U4.pic}\\ 6-U4 \end{array}$
\unitlength1mm
$\begin{array}{c} \input{7-U4.pic}\\ 7-U4 \end{array}$
\unitlength1mm
$\begin{array}{c} \input{8-U4.pic}\\ 8-U4 \end{array}$
\unitlength1mm
$\begin{array}{c} \input{9-U4.pic}\\ 9-U4 \end{array}$
\unitlength1mm
$\begin{array}{c} \input{10-U4.pic}\\ 10-U4 \end{array}$
\unitlength1mm
$\begin{array}{c} \input{11-U4.pic}\\ 11-U4 \end{array}$
\unitlength1mm
$\begin{array}{c} \input{12-U4.pic}\\ 12-U4 \end{array}$
\unitlength1mm
$\begin{array}{c} \input{13.pic}\\ 13 \end{array}$
\unitlength1mm
$\begin{array}{c} \input{14.pic}\\ 14 \end{array}$
\end{center}
\caption{
Graphs of functions $\Gamma^{(2)}$, $\Gamma^{(4)}$ in
three-loop approximation. Graphs 13 and 14 represent additional
contribution to the function $\Gamma^{(*f)}$.
In 13, 14 the $f$-vertex is indicated by a box.
}
\label{fig6}
\end{figure}

This appendix is devoted to the contributions to perturbation theory,
their representation in terms of Feynman graphs and their
corresponding loop integrals, and the evaluation of these integrals
for the two RG approaches.
Fig.\ref{fig6} shows the Feynman graphs up to third loop order
representing the contributions to the functions $\Gamma^{(2)}$ and
$\Gamma^{(4)}$ (we keep labeling of \cite{Nickel77}).

Each contribution to $\Gamma^{(*f)}$ contains the composite operator
$\prod_{i=1}^{f}\phi_{a_i}$ only once.
The relevant graphs can be obtained from the usual four-point graphs
2-U2 - 12-U4 by considering each vertex in turn to describe the
composite operator. In the three-loop approximation we consider here
two more graphs contribute which can not be
produced in this manner. They are labelled as 13 and 14.
In table \ref{tab10} we show the correspondence between the numerical
values of the loop integrals and appropriate Feynman graphs.

A diagram with $L$ loops is to be multiplied by $\sigma_d^L$ with
$$
\sigma_d= \frac{1}{2^{d-1}(\pi)^{d/2}\Gamma(d/2)},
$$
but this factor can be absorbed by redefinition of
the coupling constants $g_{ab} \rightarrow g_{ab}/\sigma_d$.

In the massless renormalization scheme loop integrals corresponding to
these graphs are evaluated by $\varepsilon$-expansion at zero mass
and non-zero external momenta chosen at the so-called symmetry point.
The expressions read \cite{Brezin76}:
\begin{eqnarray}
&&
  D_2^{\varepsilon} =  \frac{1}{\varepsilon} \Big ( 1 +
  \frac{\varepsilon}{2} + \frac{\varepsilon^2}{2} \Big),
\hspace{2.0em}
I_1^{\varepsilon} = \frac{1}{2 \varepsilon^2}
\Big ( 1 + \frac{3 \varepsilon}{2} +
\frac{5 \varepsilon^2}{2}  - \frac{J \varepsilon^2}{2} \Big ),
\nonumber \\ &&
I_2^{\varepsilon} = - \frac{1}{8\varepsilon}
\Big (1 +
\frac{5\varepsilon}{4} \Big ),
\hspace{2em}
I_3^{\varepsilon} = - \frac{1}{24\varepsilon^2} \Big (1 +
\frac{15 \varepsilon}{4} \Big ),
\nonumber \\ &&
I_4^{\varepsilon} =   \frac{1}{6 \varepsilon^3}
\Big (1 + 3 \varepsilon + \frac{31 \varepsilon^2}{4} -
\frac{3 J \varepsilon^2}{2}\Big ),
\nonumber \\ &&
I_5^{\varepsilon} =  \frac{1}{3 \varepsilon^3}
\Big (1 + \frac{5 \varepsilon}{2} +
\frac{23 \varepsilon^2}{4} - \frac{3 J \varepsilon^2}{2} \Big ),
\hspace{2em}
I_6^{\varepsilon} =   \frac{1}{3 \varepsilon^3} \Big (1 +
2 \varepsilon + \frac{13 \varepsilon^2}{4} \Big ),
\nonumber \\ &&
I_7^{\varepsilon} = \frac{\zeta(3)}{2 \varepsilon} ,
\hspace{2em}
I_8^{\varepsilon} = - \frac{1}{6 \varepsilon^2}
\Big (1 + 2 \varepsilon \Big ).
\label{a.1}
\end{eqnarray}
Here, the values of derivatives $\partial/\partial k^2$ of the
function $\Gamma^{(2)}(k)$ are given at the point $k^2=1$.
In the massive renormalization scheme loop integrals are calculated at
non-zero mass and zero external momenta (to
distinguish from (\ref{a.1}) we will label them by ``{\it m}'').
The mass renormalization introduces a higher order correction to the
propagator, which has to be taken into account in our calculation
only in the first order term (see eqs. (\ref{2.14}), (\ref{2.15})):
$$
D_2^{a_1a_2} =
D_2 + \frac{1}{9} I_2 D_{21} (u_{a_1a_1}^2 + u_{a_2a_2}^2).
$$
Here $D_{21}=(4-d)/4$. This value has been substituted into the
results for the beta functions and fixed points. $D_{21}$ does not
enter expressions which are independent of the RG scheme, such as the
resulting exponents.
The integrals
can be either $\varepsilon$-expanded (see formulas (\ref{4.11}) from
this article for instance) or numerically calculated at arbitrary
space dimensions \cite{Nickel77,Holovatch94}. In particular,
for dimension $d=2$ and $d=3$ they are given in table \ref{tab11}
with the following normalization:
$$
i_1 = I_1^m/(D_2^m)^2, \hspace{2em}  i_2 = I_2^m/(D_2^m)^2,
\hspace{2em}
i_j = I_j^m/(D_2^m)^3,  \hspace{2em} j=3 \dots 8.
$$
Note that in the massive scheme the values of the derivative
$\partial/\partial k^2$ of the
function $\Gamma^{(2)}(k)$ are given at  $k^2=0$.

\section{ Three Loop Contributions
}\label{B}
In this appendix we have collected the more lengthy expressions
for the three loop contributions to RG functions and exponents.
The coefficients $b^{jkl}$ ($j+k+l=3$) for the $\tau$
expansion of the function  $\beta_{v_{12}}^m$ (\ref{4.4}) read:
\begin{eqnarray*}
&&
b^{300}=
b^{030}=
-{\frac {{i_2}\,d}{18}}-{\frac {2\,{i_3}}{9}}-{\frac {16\,
{i_4}}{9}}-{\frac {{i_6}}{3}}-{\frac {4\,{i_8}}{9}}+{
\frac {58\,{i_1}}{27}}+{\frac {2\,{i_2}}{3}}-{\frac {20}{27}
},
      \nonumber\\ &&
b^{003}=
-{\frac {8\,{i_4}}{9}}-{\frac {2\,{i_6}}{9}}-{\frac {2\,
{i_7}}{9}}-{\frac {14}{27}}+{\frac {34\,{i_1}}{27}},
      \nonumber\\ &&
b^{210}= b^{120}= 0,
      \nonumber\\ &&
b^{102}= b^{012}=
-{\frac {20\,{i_4}}{9}}-{\frac {4\,{i_7}}{9}}-{\frac {2\,
{i_6}}{9}}-{\frac {26}{27}}+{\frac {70\,{i_1}}{27}},
      \nonumber\\ &&
b^{201}= b^{021}=
-{\frac {2\,{i_3}}{9}}-{\frac {16\,{i_4}}{9}}-{\frac {2\,
{i_7}}{3}}-{\frac {{i_2}\,d}{18}}-{\frac {28}{27}}+{\frac {20
\,{i_1}}{9}}+{\frac {2\,{i_2}}{27}},
      \nonumber\\ &&
b^{111}=
-{\frac {16}{27}}-{\frac {2\,{i_6}}{9}}-{\frac {8\,{i_4}}{9}
}+{\frac {4\,{i_1}}{3}}.
\end{eqnarray*}
The coefficients $\eta_{U;k_1;k_2}$ introduced for the $\tau$
expansion
of the exponent $\eta_{f_1f_2}^U$ in the unsymmetric fixed point U
(\ref{5.12}) read:
\begin{eqnarray*}
\eta_{U;0;{0}}&=&328-1480\,i_{{1}}-128\,i_{{2}}-240\,i_{{4}}-492\,i_
{{5}}+132\,i_{{6}}-356\,i_{{7}}-48\,i_{{8}}+2288\,{i_{{1}}}^{2}+
\\ &&
16 \,{i_{{2}}}^{2}+384\,i_{{1}}i_{{2}},
\\
\eta_{U;0;{1}}&=&-7680\,{i_{{1}}}^{2}-496\,i_{{1}}i_{{2}}+5708\,i_{{
1}}+184\,i_{{2}}+1326\,i_{{7}}+1620\,i_{{5}}-588\,i_{{6}}+204\,i_{
{4}}-
\\ &&
16\,{i_{{2}}}^{2}+48\,i_{{8}}-1312 ,
\\
\eta_{U;0;{2}}&=&570+810\,i_{{4}}+1488\,{i_{{1}}}^{2}-24\,i_{{2}}+48
\,i_{{1}}i_{{2}}-2154\,i_{{1}}-270\,i_{{7}}+216\,i_{{6}}-216\,i_{{
5}} ,
\\
\eta_{U;0;{3}}&=&-54-162\,i_{{4}}+162\,i_{{1}} ,
\\
\eta_{U;1;{0}}&=&-16\,{i_{{2}}}^{2}+468\,i_{{7}}-448\,i_{{1}}i_{{2}}
+160\,i_{{2}}+2408\,i_{{1}}-252\,i_{{6}}-2992\,{i_{{1}}}^{2}+564\,
i_{{5}}+
\\ &&
48\,i_{{8}}-560 ,
\\
\eta_{U;1;{1}}&=&-5342\,i_{{1}}+176\,i_{{1}}i_{{2}}-88\,i_{{2}}+756
\,i_{{6}}-594\,i_{{7}}-252\,i_{{5}}+3536\,{i_{{1}}}^{2}+1638\,i_{{
4}}+1346 ,
\\
\eta_{U;1;{2}}&=&1188\,i_{{1}}-216\,i_{{5}}-216\,i_{{6}}-756\,i_{{4}
}-324 ,
\\
\eta_{U;2;{0}}&=&-1072\,i_{{1}}-32\,i_{{2}}+64\,i_{{1}}i_{{2}}+336\,
i_{{4}}-112\,i_{{7}}+144\,i_{{6}}+704\,{i_{{1}}}^{2}-48\,i_{{5}}+
272 ,
\\
\eta_{U;2;{1}}&=&1098\,i_{{1}}-180\,i_{{6}}-738\,i_{{4}}-180\,i_{{5}
}-306 ,
\\
\eta_{U;3;{0}}&=&-40+144\,i_{{1}}-24\,i_{{6}}-96\,i_{{4}}-24\,i_{{5}
}.
\end{eqnarray*}
\end{appendix}

\bibliographystyle{unsrt}

\begin{thebibliography}{80}
\bibitem{note3}
Note that our three loop result (\ref{4.3}) is more general than the
corresponding expression for the $d=3$ $\beta$-function in
\cite{Nickel78}. It gives the dimensional dependence of the $\beta$-function
and can be evaluated for arbitrary fixed $d$.
\bibitem{note5}
This value for the two-loop approximation was obtained in
  \cite{Holovatch93a}. Due to different normalizations
  $u^*$(Ref.\cite{Holovatch93a})$=\frac{4}{3}v_{aa}^*$(here)$=1.5809$. In three
  loop approximation the corresponding value equals $u^*=1.4393$
  \cite{Holovatch93a}, in six loops one has $u^*=1.421 \pm 0.004$
  \cite{Nickel78}.
\bibitem{note6}
The system of equations for the fixed point is solved at given order
of the perturbation theory. With increasing order the difference
between the results should decrease.
\bibitem{note7}
Here the appropriate value of the resummation parameter is $a=a^S$.
\end{thebibliography}


\begin{table}[H]
\caption { \label{tab3}
Values of the copolymer star exponent $\eta_{f_1f_2}$ obtained in
the first, second, and third order in the
Gaussian ($G$) fixed point in $\varepsilon$-expansion and
pseudo-$\varepsilon$ ($\tau$) expansion for different
values of $f_1, f_2$ at $\varepsilon=1$ ($d=3$).
$res$ stands for the
results obtained by Pad\'e-Borel resummation of
the three loop series.}
\begin{centering}
 \begin{tabular}{rrrrrrrrrr}
$f_1$ & $f_2$ & $\sim \varepsilon$ & $\sim \varepsilon^2$   &
$\sim \varepsilon^3$ & $res.$ & $\sim \tau$ & $\sim \tau^2$ &
$\sim \tau^3$ & $res.$  \\
\hline
1 & 1 & -.50 & -.63 & -.46 & -.56  & -.50&  -.58&  -.56&  -.57  \\
1 & 2 & -1.00& -1.00& -1.00&       & -1.00& -1.00& -1.00&       \\
1 & 3 & -1.50& -1.13& -1.99& -1.36 & -1.50& -1.25& -1.42& -1.34 \\
2 & 2 & -2.00& -1.50& -2.65& -1.81 & -2.00& -1.67& -1.93& -1.80 \\
2 & 3 & -3.00& -1.50& -5.71& -2.50 & -3.00& -2.00& -3.01& -2.45 \\
3 & 3 & -4.50& -1.13& -12.27&-3.48 & -4.50& -2.25& -5.09& -3.37 \\
 \end{tabular}
\end{centering}
 \end{table}

\begin{table}[H]
\caption {
\label{tab4}
Values of the copolymer star exponent $\eta_{f_1f_2}$ obtained in
the first, second, and third order in the
unsymmetrical ($U$) fixed point in $\varepsilon$-expansion and
pseudo-$\varepsilon$ ($\tau$) expansion for different
values of $f_1, f_2$ at $\varepsilon=1$ ($d=3$).
$res$ stands for the
results obtained by Pad\'e-Borel resummation of
the three loop series.}
\begin{centering}
 \begin{tabular}{rrrrrrrrrr}
$f_1$ & $f_2$ & $\sim \varepsilon$ & $\sim \varepsilon^2$   &
$\sim \varepsilon^3$ & $res.$ & $\sim \tau$ & $\sim \tau^2$ &
$\sim \tau^3$ & $res.$  \\
\hline
1& 1& -.38 & -.50 & -.28&   -.43 &  -.38&  -.46&  -.43&  -.44  \\
1& 2& -.75 &  -.85& -.69&   -.80 &  -.75&  -.82&  -.78&  -.80  \\
1& 3& -1.13& -1.07&-1.33&  -1.11 & -1.13& -1.09& -1.11& -1.10  \\
2& 1& -1.00& -.98 & -.71& -1.00  & -1.00&  -.99&  -.98& -.99   \\
2& 2& -1.75& -1.37& -2.37&  -1.62& -1.75& -1.50& -1.71& -1.60  \\
2& 3& -2.50& -1.47& -4.99&  -2.19& -2.50& -1.82& -2.56& -2.13  \\
3& 1& -1.88& -1.28& -1.70&  -1.50& -1.88& -1.48& -1.82& -1.64  \\
3& 2& -3.00& -1.36& -6.19&  -2.47& -3.00& -1.91& -3.18& -2.43  \\
3& 3& -4.13& -1.02&-12.83&  -3.26& -4.13& -2.06& -4.97& -3.14  \\
 \end{tabular}
\end{centering}
 \end{table}

\begin{table}[H]
\caption{ \label{tab5}
Values of the copolymer star exponent $\eta_{f_1f_2}^{G}$
at $d=3$ obtained by $\varepsilon$-expansion
($\varepsilon$) and by fixed dimension technique
($3d$).
}
\tabcolsep1.4mm
\begin{tabular}{lrrrrrrrrrrrr}
$f_1$ &
\multicolumn{2}{c}{$1$}&
\multicolumn{2}{c}{$2$}&
\multicolumn{2}{c}{$3$}&
\multicolumn{2}{c}{$4$}&
\multicolumn{2}{c}{$5$}&
\multicolumn{2}{c}{$6$}\\
$f_2$ &
$\varepsilon$ & $3d$ &
$\varepsilon$ & $3d$ &
$\varepsilon$ & $3d$ &
$\varepsilon$ & $3d$ &
$\varepsilon$ & $3d$ &
$\varepsilon$ & $3d$ \\
\hline
1 &
  -0.56 &  -0.58 &
  -1.00 &  -1.00 &
  -1.33 &  -1.35 &
  -1.63 &  -1.69 &
  -1.88 &  -1.98 &
  -2.10 &  -2.24 \\
2 & & &
  -1.77 &  -1.81 &
  -2.45 &  -2.53 &
  -3.01 &  -3.17 &
  -3.51 &  -3.75 &
  -3.95 &  -4.28 \\
3 & & & & &
 -3.38 &  -3.57 &
 -4.21 &  -4.50 &
 -4.94 &  -5.36 &
 -5.62 &  -6.15 \\
4 & & & & & & &
 -5.27 &  -5.71 &
 -6.24 &  -6.84 &
 -7.12 &  -7.90 \\
5 & & & & & & & & &
 -7.42 &  -8.24 &
 -8.50 &  -9.54 \\
6 & & & & & & & & & & &
 -9.78 &  -11.07
\end{tabular}
\end{table}

\begin{table}[H]
\caption{ \label{tab6}
Values of the copolymer star exponent $\eta_{f_1f_2}^{U}$
at $d=3$ obtained by $\varepsilon$-expansion
($\varepsilon$) and by fixed dimension technique
($3d$).
}
\tabcolsep1.4mm
\begin{tabular}{lrrrrrrrrrrrrr}
$f_1$ &
\multicolumn{2}{c}{$1$}&
\multicolumn{2}{c}{$2$}&
\multicolumn{2}{c}{$3$}&
\multicolumn{2}{c}{$4$}&
\multicolumn{2}{c}{$5$}&
\multicolumn{2}{c}{$6$}\\
$f_2$ &
$\varepsilon$ & $3d$ &
$\varepsilon$ & $3d$ &
$\varepsilon$ & $3d$ &
$\varepsilon$ & $3d$ &
$\varepsilon$ & $3d$ &
$\varepsilon$ & $3d$ \\
\hline
0 &
    0 &  0    &
-0.28 & -0.28 &
-0.75 & -0.76 &
-1.36 & -1.38 &
-2.07 & -2.14 &
-2.88 & -3.01 \\
1 &
-0.43 &  -0.45 &
-0.98 &  -0.98 &
-1.64 &  -1.67 &
-2.39 &  -2.47 &
-3.21 &  -3.38 &
-4.11 &  -4.40 \\
2 &
-0.79 &  -0.81 &
-1.58 &  -1.60 &
-2.44 &  -2.52 &
-3.33 &  -3.50 &
-4.28 &  -4.57 &
-5.29 &  -5.73 \\
3 &
-1.09 &  -1.09 &
-2.13 &  -2.19 &
-3.16 &  -3.30 &
-4.20 &  -4.48 &
-5.28 &  -5.71 &
-6.41 &  -7.03 \\
4 &
-1.35 &  -1.37 &
-2.61 &  -2.71 &
-3.82 &  -4.04 &
-5.02 &  -5.40 &
-6.24 &  -6.81 &
-7.48 &  -8.28 \\
5 &
-1.60 &  -1.64 &
-3.05 &  -3.21 &
-4.44 &  -4.75 &
-5.80 &  -6.30 &
-7.15 &  -7.89 &
-8.51 &  -9.50 \\
6 &
-1.81 &  -1.89 &
-3.46 &  -3.68 &
-5.01 &  -5.42 &
-6.53 &  -7.15 &
-8.02 &  -8.92 &
-9.50 & -10.69
\end{tabular}
\end{table}

\begin{table}[H]
\caption{ \label{tab7}
Values of $\eta^{MAW}_f$ exponents of star of
mutually avoiding walks
at $d=3$, $d=2$ obtained by  $\varepsilon$-expansion
($\varepsilon$) and by fixed dimension technique
($3d$, $2d$).
The last column gives the exact conjecture
at $d=2$
\protect\cite{Duplantier88,Duplantier88a}.
}
\begin{tabular}{llllll}
&\multicolumn{2}{c} {$d=3$}&\multicolumn{3}{c} {$d=2$} \\
$f$ & $\varepsilon$ & $3d$ &
$\varepsilon$ & $2d$
& $exact$ \\
\hline
1  &  0     &   0    &   0     &   0     & -.250      \\
2  & -.56   &  -.56  & -1.20   & -1.19   & -1.250     \\
3  & -1.38  &  -1.36 & -2.71   & -2.60   & -2.916(6)  \\
4  & -2.36  &  -2.34 & -4.36   & -4.07   & -5.250     \\
5  & -3.43  &  -3.43 & -6.04   & -5.61   & -8.250     \\
6  & -4.58  &  -4.64 & -7.78   & -7.17   & -11.916(6)
\end{tabular}
\end{table}

\begin{table}[H]
\caption{ \label{tab8}
Values of the copolymer star exponent $\eta_{f_1f_2}^{G}$
at $d=2$ obtained by $\varepsilon$-expansion
($\varepsilon$) and by fixed dimension technique
($2d$).
}
\tabcolsep1.4mm
\begin{tabular}{lrrrrrrrrrrrr}
$f_1$ &
\multicolumn{2}{c}{$1$}&
\multicolumn{2}{c}{$2$}&
\multicolumn{2}{c}{$3$}&
\multicolumn{2}{c}{$4$}&
\multicolumn{2}{c}{$5$}&
\multicolumn{2}{c}{$6$}\\
$f_2$ &
$\varepsilon$ & $2d$ &
$\varepsilon$ & $2d$ &
$\varepsilon$ & $2d$ &
$\varepsilon$ & $2d$ &
$\varepsilon$ & $2d$ &
$\varepsilon$ & $2d$ \\
\hline
1 &
 -1.20  & -1.22  &
 -1.98  & -1.98  &
 -2.56  & -2.58  &
 -2.99  & -3.04  &
 -3.36  & -3.43  &
 -3.68  & -3.78  \\
2 & & &
-3.41  & -3.45  &
-4.49  & -4.59  &
-5.37  & -5.52  &
-6.13  & -6.34  &
-6.80  & -7.04  \\
3 & & & & &
-6.05 &  -6.23 &
-7.36 &  -7.63 &
-8.49 &  -8.84 &
-9.50 &  -9.91 \\
4 & & & & & & &
-9.06 &  -9.44   &
-10.55 &  -11.03  &
-11.89 &  -12.45  \\
5 & & & & & & & & &
-12.38 &  -12.98 &
-14.03 &  -14.74 \\
6 & & & & & & & & & & &
-15.99 &  -16.81
\end{tabular}
\end{table}

\begin{table}[H]
\caption{ \label{tab9}
Values of the copolymer star exponent $\eta_{f_1f_2}^{U}$
at $d=2$ obtained by $\varepsilon$-expansion
($\varepsilon$) and by fixed dimension technique
($2d$).
}
\tabcolsep1.4mm
\begin{tabular}{lrrrrrrrrrrrr}
$f_1$ &
\multicolumn{2}{c}{$1$}&
\multicolumn{2}{c}{$2$}&
\multicolumn{2}{c}{$3$}&
\multicolumn{2}{c}{$4$}&
\multicolumn{2}{c}{$5$}&
\multicolumn{2}{c}{$6$}\\
$f_2$ &
$\varepsilon$ & $2d$ &
$\varepsilon$ & $2d$ &
$\varepsilon$ & $2d$ &
$\varepsilon$ & $2d$ &
$\varepsilon$ & $2d$ &
$\varepsilon$ & $2d$ \\
\hline
0 &
  0   &  0     &
-0.59 & -0.62  &
-1.51 & -1.53  &
-2.61 &  -2.63 &
-3.84 &  -3.89 &
-5.18 &   -5.28 \\
1 &
-.91  & -.96   &
-1.94 & -1.96  &
-3.11 & -3.13  &
-4.35 &  -4.41 &
-5.71 &  -5.83 &
-7.17 &  -7.35 \\
2 &
-1.62 & -1.63  &
-3.05 & -3.09  &
-4.49 & -4.54  &
-5.94 &  -6.06 &
-7.46 &  -7.64 &
-9.04 &  -9.30 \\
3 &
 -2.16 & -2.16  &
 -4.00 & -4.04  &
 -5.70 & -5.80  &
 -7.39 &  -7.57 &
 -9.09 &  -9.35 &
-10.82 &  -11.17\\
4 &
 -2.60 & -2.64  &
 -4.80 & -4.88  &
 -6.79 & -6.96  &
 -8.72 &  -8.97 &
-10.61 &  -10.95&
-12.52 &  -12.95\\
5 &
 -3.00 & -3.03  &
 -5.52 & -5.63  &
 -7.81 & -8.01  &
 -9.96 &  -10.27&
-12.06 &  -12.47&
-14.14 &  -14.65\\
6 &
 -3.34 & -3.39  &
 -6.17 & -6.33  &
 -8.73 & -8.99  &
-11.12 &  -11.49&
-13.43 &  -13.91&
-15.69 &  -16.27
\end{tabular}
\end{table}

\begin{table} [H]
\caption { \label{tab10}
Values of the loop integrals. The graphs 2-M1 - 5-S3 indicate the
derivative of the two-point function $\partial/\partial k^{2}
\Gamma^{(2)}(k)$.}
\begin{centering}
\begin{tabular}{llllll}
\\
Graph&Integral&Graph&Integral&Graph&Integral  \\
 &value& &value& &value  \\
 \hline
\\
2-U2 & $D_2$     & 8-U4  &  $I_4$ &  14   &  $I_1D_2$ \\
3-U3 & $D_2^2$     & 9-U4  &  $I_5$ &  2-M1 &  0    \\
4-U3 & $I_1$ & 10-U  &  $I_6$ &  3-S2 &  $I_2$ \\
5-U4 & $D_2^3$     & 11-U4 &  $I_5$ &  4-M3 &  0    \\
6-U4 & $I_1D_2$ & 12-U4 &  $I_7$ &  5-S3 &  $I_8$ \\
7-U4 & $I_3$ & 13    &  $D_2^3$     &       &        \\
\end{tabular}
\end{centering}
\end{table}

\begin{table}[H]
\caption { \label{tab11}
Numerical values of normalized loop integrals $i_j$ calculated
in the massive field theory framework
\protect\cite{Nickel77,Holovatch94}.
}
\begin{centering}
\begin{tabular}{lllll}
&$i_1$& $i_2$& $i_3$& $i_4$ \\
$d=2$ & 0.781302412896 & -0.114635746230 & -0.044703881514 &
0.569829439192 \\
$d=3$ & 0.6666666667 & -0.0740740741 & -0.0376820725 &
0.3835760966 \\
\hline
& $i_5$& $i_6$& $i_7$& $i_8$ \\
$d=2$ & 0.659043562065 & 0.650899895132 &
0.40068563 & -0.157398409771 \\
$d=3$ & 0.5194312413 & 0.5000000000 &
0.1739006107 & -0.0946514319
\end{tabular}
\end{centering}
\end{table}

\end{document}